\documentclass[sigconf, nonacm]{acmart}

\usepackage{subfiles}
\usepackage{makecell}
\usepackage{enumitem}
\usepackage{multirow}
\usepackage{longtable}
\usepackage{subfigure}
\usepackage{graphicx}
\usepackage{caption}
\usepackage{subcaption}

\usepackage{enumitem}
\setlist{nolistsep}

\newcommand{\dquote}[1]{\textit{``#1''}}

\newcommand{\groupone}{\textsc{\textbf{G1}}}
\newcommand{\grouptwo}{\textsc{\textbf{G2}}}
\newcommand{\groupthree}{\textsc{\textbf{G3}}}
\newcommand{\groupfour}{\textsc{\textbf{G4}}}
\newcommand{\groupfive}{\textsc{\textbf{G5}}}
\newcommand{\grouponeunstyled}{G1}
\newcommand{\grouptwounstyled}{G2}
\newcommand{\groupthreeunstyled}{G3}
\newcommand{\groupfourunstyled}{G4}
\newcommand{\groupfiveunstyled}{G5}
\newcommand{\cattlehug}{\textbf{\textit{\grouponeunstyled{}D}}}
\newcommand{\plotsnail}{\textbf{\textit{\grouponeunstyled{}M}}}
\newcommand{\sunnygoat}{\textbf{\textit{\grouponeunstyled{}U1}}}
\newcommand{\pigwhisper}{\textbf{\textit{\grouponeunstyled{}U2}}}
\newcommand{\tunakept}{\textbf{\textit{\grouponeunstyled{}U3}}}
\newcommand{\floorlight}{\textbf{\textit{\grouptwounstyled{}D}}}
\newcommand{\stewmeter}{\textbf{\textit{\grouptwounstyled{}M}}}
\newcommand{\jewelpipe}{\textbf{\textit{\grouptwounstyled{}U1}}}
\newcommand{\glidemail}{\textbf{\textit{\grouptwounstyled{}U2}}}
\newcommand{\tribeboat}{\textbf{\textit{\grouptwounstyled{}U3}}}
\newcommand{\ironbush}{\textbf{\textit{\groupthreeunstyled{}D}}}
\newcommand{\subtlerun}{\textbf{\textit{\groupthreeunstyled{}M}}}
\newcommand{\hoodclad}{\textbf{\textit{\groupthreeunstyled{}U1}}}
\newcommand{\lemonfret}{\textbf{\textit{\groupthreeunstyled{}U2}}}
\newcommand{\bootspark}{\textbf{\textit{\groupthreeunstyled{}U3}}}
\newcommand{\mailchart}{\textbf{\textit{\groupfourunstyled{}D}}}
\newcommand{\shellhusk}{\textbf{\textit{\groupfourunstyled{}M}}}
\newcommand{\brickdew}{\textbf{\textit{\groupfourunstyled{}U1}}}
\newcommand{\onlybelt}{\textbf{\textit{\groupfourunstyled{}U2}}}
\newcommand{\hollowsea}{\textbf{\textit{\groupfourunstyled{}U3}}}
\newcommand{\stampjab}{\textbf{\textit{\groupfiveunstyled{}D}}}
\newcommand{\webcube}{\textbf{\textit{\groupfiveunstyled{}M}}}
\newcommand{\gelpick}{\textbf{\textit{\groupfiveunstyled{}U1}}}
\newcommand{\hallpaste}{\textbf{\textit{\groupfiveunstyled{}U2}}}
\newcommand{\midfrost}{\textbf{\textit{\groupfiveunstyled{}U3}}}

\newif\ifredact
\redactfalse
\newcommand{\redact}[2]{\ifredact #1\else #2\fi}

\AtBeginDocument{
  }

\begin{document}

\title{Sword and Shield: Uses and Strategies of LLMs in Navigating Disinformation}

\author{Gionnieve Lim}
\email{gionnievelim@gmail.com}
\orcid{0000-0002-8399-1633}
\affiliation{
    \institution{Singapore University of Technology and Design}
    \country{Singapore}
}

\author{Bryan Chen Zhengyu Tan}
\email{bryantanzy@gmail.com}
\orcid{0009-0002-0428-9641}
\affiliation{
  \institution{Singapore University of Technology and Design}
  \country{Singapore}
}

\author{Kellie Yu Hui Sim}
\email{kelliesyhh@gmail.com}
\orcid{0009-0005-6451-7089}
\affiliation{
  \institution{Singapore University of Technology and Design}
  \country{Singapore}
}

\author{Weiyan Shi}
\email{weiyanshi6@gmail.com}
\orcid{0009-0001-6035-9678}
\affiliation{
  \institution{Singapore University of Technology and Design}
  \country{Singapore}
}

\author{Ming Hui Chew}
\email{chewminghuis@gmail.com}
\orcid{0009-0003-1497-4287}
\affiliation{
  \institution{Singapore University of Technology and Design}
  \country{Singapore}
}

\author{Ming Shan Hee}
\email{mingshan_hee@mymail.sutd.edu.sg}
\orcid{0000-0002-6328-5889}
\affiliation{
  \institution{Singapore University of Technology and Design}
  \country{Singapore}
}

\author{Roy Ka-Wei Lee}
\email{roy_lee@sutd.edu.sg}
\orcid{0000-0002-1986-7750}
\affiliation{
  \institution{Singapore University of Technology and Design}
  \country{Singapore}
}

\author{Simon T. Perrault}
\email{perrault.simon@gmail.com}
\orcid{0000-0002-3105-9350}
\affiliation{
    \institution{Singapore University of Technology and Design}
    \country{Singapore}
}

\author{Kenny Tsu Wei Choo}
\email{kennytwchoo@gmail.com}
\orcid{0000-0003-3845-9143}
\affiliation{
  \institution{Singapore University of Technology and Design}
  \country{Singapore}
}

\renewcommand{\shortauthors}{Lim et al.}

\begin{abstract}
The emergence of Large Language Models (LLMs) presents a dual challenge in the fight against disinformation. These powerful tools, capable of generating human-like text at scale, can be weaponised to produce sophisticated and persuasive disinformation, yet they also hold promise for enhancing detection and mitigation strategies. This paper investigates the complex dynamics between LLMs and disinformation through a communication game that simulates online forums, inspired by the game \textit{Werewolf}, with 25 participants. We analyse how Disinformers, Moderators, and Users leverage LLMs to advance their goals, revealing both the potential for misuse and combating disinformation. Our findings highlight the varying uses of LLMs depending on the participants' roles and strategies, underscoring the importance of understanding their effectiveness in this context. We conclude by discussing implications for future LLM development and online platform design, advocating for a balanced approach that empowers users and fosters trust while mitigating the risks of LLM-assisted disinformation.

\end{abstract}




\maketitle

\section{Introduction}

The advent of Large Language Models (LLMs) such as GPT-4o\footnote{https://openai.com/index/hello-gpt-4o/} and Llama 3\footnote{https://ai.meta.com/blog/meta-llama-3/} have brought about a profound shift in digital communication, with significant implications for the spread of information and disinformation on social media platforms.
These models can generate human-like text at an unprecedented scale and speed and potentially transform both the creation and dissemination of content online.
While LLMs offer exciting possibilities for enhancing communication and automation, they also pose new risks by enabling the rapid and widespread generation of persuasive disinformation despite the increasing focus and development of safeguards~\cite{dong2024safeguarding}.

Social media, already a fertile ground for the spread of disinformation~\cite{geeng_fakenews_fbtwt_2020}, has been further complicated by the rise of LLMs.
As put by DiResta, \dquote{social media took the cost of distribution to zero, and generative AI takes the cost of generation to zero}~\cite{ProducingFakeInformation}.
These models can be used to generate content that mimics the tone, style, and nuances of human communication, making it increasingly difficult for users to discern between genuine and fabricated information~\cite{chen2023can, spitale2023ai}. More concerning, false and misleading claims that are contextually relevant to significant societal topics can be easily produced~\cite{brewster2023could}.
Disinformation can range from small, localised settings~\cite{KARELL2022101641,publications10020015} to large-scale campaigns~\cite{HeresWhatWe, benkler2018network}, and such attempts are already underway.
NewsGuard has identified 1065 unreliable AI-generated news websites that span 16 languages~\cite{TrackingAIenabledMisinformation}. OpenAI has reported foiling five influence operations that used their models to generate, translate and convert content to post on multiple platforms~\cite{DisruptingDeceptiveUses}. Yet, the same tech also shows promise to mitigate disinformation, being incorporated for detecting harmful content~\cite{GenerativeAIAlready} and fact-checking~\cite{ThisBrazilianFactchecking}.
As LLMs become more integrated into the fabric of online communication, their role in shaping public discourse--both positively and negatively--cannot be overstated.

Many works illuminate the scope and mechanisms of disinformation, but as access to real-world disinformers is difficult~\cite{IndexonCensorship2011, Nimmo2018}, they often rely on indirect methods such as accounts from the mitigators~\cite{mirza2023tactics}, text analyses~\cite{Vosoughi2018, Recuero_Soares_Vinhas_2020, KING_PAN_ROBERTS_2013}, algorithmic signals~\cite{hussain2018analyzing, Vargas2020, Alieva2022}, and case studies~\cite{10.1145/3359229, 10.1145/3579616}.
While these approaches reveal insights on disinformation, they face challenges in conclusively inferring the intent to deceive and may fail to capture the adaptive strategies that disinformers employ; a situation made more complex with LLMs joining the scene~\cite{chen2023combating}.
In contrast, a simulated game environment enables direct observation of disinformers’ decision-making processes, offering a controlled yet dynamic space to study how LLMs are leveraged to fabricate narratives, tailor content, and adapt strategies. While also an indirect method, this complementary approach yields valuable exploratory insights into disinformation and countermeasures in emerging scenarios involving LLMs, where the actions and intents of various actors can be reliably captured.

In this paper, we explore the potential impact of LLMs on the dynamics of disinformation in small-scale intimate settings through a communication game inspired by the classic game of \textit{Werewolf}. 
Participants interact within our controlled online environment, taking on the roles of Disinformers (Werewolves), Moderators (Seers), and regular Users (Villagers) and participating in a simulated online forum. 
Each participant is equipped with an LLM to assist them in either spreading or countering disinformation, and either to avoid being detected as a Disinformer or to identify them.
By incorporating an LLM into this simulation, we explore its impact on the strategies used by Disinformers, Moderators, and Users. 

We map out the potential ways in which LLMs
offer new avenues for disinforming as well as its detection and mitigation.
Notably, we reveal how LLMs are not only used for informational purposes like creating false and persuasive content or assisting in the verification of content, uses that have been predominantly perceived to be the key advantages that LLMs bring to the disinformation landscape~\cite{chen2023combating, shah2024navigating}, but also serve strategic advisory purposes for all actors, both malicious and virtuous, supporting them to plan actions aligned with their respective goals.

Our findings offer insights into the rapidly evolving technologies of LLMs in the disinformation scene, how they interact with users, and the social influences on digital platforms, shedding light on their potential impact on the future of online communication.
Specifically, this work contributes the following:
\begin{itemize}
    \item We present a pioneering study exploring how online community actors engage with disinformation through a communication game. This study uniquely examines the unfolding of disinformation strategies across different roles—\allowbreak Disinformers, Moderators, and Users—and how LLMs influence these dynamics.
    \item We investigate the capabilities of LLMs as tools for amplifying and countering disinformation by analysing how ill/well-intentioned actors use them to either generate and refine disinformation or detect and mitigate it. Our study also uncovers the varying efficacy of LLMs depending on user roles and strategies.
    \item We examine broader design and policy implications for LLM-assisted disinformation detection and mitigation. We also propose solutions for enhancing user interaction with LLMs, fostering appropriate reliance, and integrating AI systems into social platforms to balance disinformation combative efforts and human oversight.
\end{itemize}

\section{Related Work}

\subsection{Disinformation in Online Platforms}

Disinformation refers to the \textit{intentional} creation and dissemination of false content to deceive or manipulate others~\cite{Wardle2017}. Persuasion techniques are often employed in disinformation to influence public opinion and behaviour~\cite{guess2020misinformation}. As such, we consider disinformation as content that was created with the intention to manipulate opinions through falsehoods and persuasive strategies in this work.
Disinformation can manifest as fabricated content~\cite{CommonsLibrarian2024}, conspiracy theories~\cite{10.1145/3610043}, and rumours~\cite{antjeDisinformationCommonForms2024}. Motivations range from profit-driven clickbait~\cite{scott2023deceptive} and trolling~\cite{10.1145/3449215} to advancing ideological agendas~\cite{keeley2024conspiracy}. Disinformers, including individuals, activist groups, businesses, and state actors~\cite{mirza2023tactics}, pursue outcomes like financial gain~\cite{Grohmann2024}, reputational benefits~\cite{deJong2020} and power~\cite{HeresWhatWe}. They work to amplify narratives~\cite{Shao2018} or engage deeply to manipulate emotions like anger and fear~\cite{Sunstein2009}.
Tactics include crafting convincing profiles, seeding emotional narratives, and impersonating authoritative figures to establish credibility~\cite{mirza2023tactics, KING_PAN_ROBERTS_2013}. Persuasion techniques like emotional framing, urgency, and selective evidence are also often exploited~\cite{Recuero_Soares_Vinhas_2020, CHEN2021}. Social network analyses have revealed coordinated propagation behaviours~\cite{hussain2018analyzing, Vargas2020, Alieva2022}.

Disinformation is often associated with large-scale, coordinated campaigns, but it also proliferates through intimate, interpersonal interactions within social media. In these environments, individual users may share misleading or false information through casual posts, personal anecdotes, or opinion-based commentary~\cite{Buchanan2020}. 
These localised incidents can be highly impactful, influencing community-level opinions~\cite{Potter2021} and deepening social divides~\cite{VanRaemdonck2019}. Comments with false or misleading information further receive replies that contain falsehoods or are inflammatory~\cite{Buchanan2022}. These patterns highlight how disinformation can take root within online communities, amplifying polarization.
The rise of LLMs complicates detection, as these models generate realistic and persuasive content that can fuel disinformation~\cite{meier2024llm}. While LLMs hold potential for fact-checking and moderation, their misuse for fabricating narratives presents a significant challenge. This study aims to advance the understanding of their dual role in spreading and countering disinformation by examining the impacts of LLMs on disinformation practices in social media, particularly at small and localised settings.

\subsection{Disinformation in the Age of LLMs}
Advancements in LLMs have sparked concerns about their potential to amplify disinformation~\cite{goldstein2023generative}. 
While LLM providers continuously reinforce safeguards to prevent misuse~\cite{OpenAISafetyPractices, TrustSafety}, the proliferation of \textit{uncensored}, \textit{open-source} LLMs online makes it a persistent challenge. 
The simplicity of using these models enables the rapid creation of vast amounts of varied misleading content~\cite{barman2024dark}. 
The ability of LLMs to produce context-sensitive, coherent, and persuasive responses that are hard to discern from human-generated content further heightens the risk of disinformation~\cite{spitale2023ai}.

Since the launch of ChatGPT~\cite{LargeLanguageModel}, LLM adoption in both commercial and personal contexts has surged, prompting a parallel growth in research focused on detecting AI-generated content. 
While numerous techniques have been developed, their success varies~\cite{wu2023survey}. 
A significant challenge is that LLM-generated content often mimics credible evidence, cites sources, and acknowledges limitations, making it harder for detection models—and humans—to identify falsehoods~\cite{zhou2023synthetic}. 
Research shows that individuals generally struggle to differentiate between human- and machine-generated content~\cite{chen2023can, spitale2023ai}. 
Kreps et al.~\cite{kreps2022all} noted that both content are perceived as equally credible, raising concerns about how malicious actors could exploit this trust to erode confidence in democratic institutions.

Despite their potential to spread disinformation, LLMs also offer tools to combat it. This dual nature has been described as a \dquote{double-edged sword}~\cite{shah2024navigating}. 
With vast knowledge and strong reasoning capabilities, LLMs can be harnessed to detect disinformation across a wide range of topics~\cite{chen2023combating}, though their effectiveness depends on the prompts and models used~\cite{chen2023can, pelrine2023towards}.
Being able to detect factual claims~\cite{ni2024afacta, choi2024automated}, generate relevant web queries~\cite{setty2024factcheck}, summarise content, and provide reasoning~\cite{hsu2024enhancing, quelle2024perils}, LLMs can significantly automate the fact-checking processes, allowing information to be verified and disseminated more efficiently. 
As LLMs become ubiquitous for everyday information retrieval~\cite{skjuve2024people}, it is likely that individuals will also use them to assess content and guard against falsehoods. 
While prior research has explored LLMs in various everyday information contexts, our focus is on how LLMs can assist regular users in identifying bad actors and countering disinformation.

\subsection{Disinformation Detection and Mitigation}

The rise of disinformation has prompted diverse detection and mitigation strategies, ranging from algorithmic to human-centric approaches. Algorithmic approaches have shown promise in scalability and accuracy~\cite{Zhou2020, Wu2019, Islam2020, Mridha2021}. However, their application on social media has been primarily limited to content downranking or removal~\cite{Saltz2021b}. Human oversight, especially through fact-checking labels, remains a key countermeasure despite scalability challenges~\cite{Uscinski2013, Moy2021}. Alternatives like crowdsourcing and automated measures have shown potential, with studies indicating that fact-checking labels reduce engagement with false content~\cite{Seo2019, Lanius2021, Jia2022}.

Moreover, recent advancements have seen the integration of LLMs into various stages of the disinformation detection and mitigation pipeline. LLMs have been employed as social knowledge providers~\cite{pavlyshenkoAnalysisDisinformationFake2023, laiRumorLLMRumorLarge2024, liuTELLERTrustworthyFramework2024}, external helpers~\cite{liSelfcheckerPlugplayModules2024, panFactcheckingComplexClaims2023}, and decision makers~\cite{wanGeneratingReactionsExplanations2024, yangReinforcementTuningDetecting2024} in disinformation detection systems. These applications leverage the expansive knowledge base and reasoning capabilities of LLMs to enhance the performance of automated fact-checking systems.

Despite these advancements, human judgement remains crucial for nuanced interpretation, ethical considerations, and adapting to evolving disinformation tactics~\cite{micallef_factcheck_2022}. Our research aims to address a critical gap in this landscape by examining how everyday users, not just professionals, can leverage LLMs in disinformation mitigation. By simulating an online forum environment, we seek to inform the design of more effective and accessible disinformation mitigation strategies that are aligned with the needs and behaviours of everyday social media users, as well as offer unique insights into the interplay between LLMs and different actors within the prevalent context of online social forums.

\subsection{Communication Games}
Communication games offer an economical and efficient way to simulate complex social interactions, enabling the study of individual and group dynamics in controlled settings~\cite{verhagen2017games}. Despite their limitations in replicating real-world scenarios~\cite{Lukosch2018}, participants' attitudes, behaviours, and intentions can reflect characteristics of their real-world counterparts in specific interactive scenarios~\cite{white2018communication}, giving insights into social phenomena often inaccessible to researchers~\cite{Vemuri2020}.

The \textit{Werewolf} game has been considered a valuable tool in communication and reasoning research, applied to teaching communication skills~\cite{tilton2019winning}, assessing social interactions~\cite{zhang2021sense}, and examining LLMs' reasoning capabilities~\cite{tsunoda2019ai, toriumi2017ai, nakamura2016constructing}. \textit{Werewolf} requires players to identify hidden roles in a situation of asymmetric information, and the social deduction process makes it ideal for studying patterns of information exchange.

The emphasis on communication, deduction, and strategy makes \textit{Werewolf} a fitting model to examine how users interact with each other and LLMs in contexts of information exchange and disinformation detection. Our study adopts its game mechanics to simulate interactions among disinformers, moderators, and users on social media in the presence of LLMs, analysing how LLMs influence disinformation and counteractive behaviours. 

\section{Method}

\subsection{LLM-powered Werewolf Communication Game}
We designed and developed a communication game inspired by the classic \textit{Werewolf} game~\cite{xu_exploring_2023} to simulate an online forum--similar to Reddit--where a Disinformer attempts to influence the community, and empowered all roles with our built-in chatbot. 
\textit{Werewolf} offers several parallels to disinformation on social media:

\begin{itemize}
    \item A bad actor (Werewolf/Disinformer) hidden among common actors (Villagers/Users)
    \item A good actor empowered to assist the community (Seer/\allowbreak Moderator)
    \item Use of deception and misdirection by the Disinformer to avoid detection
    \item The community's goal is to identify and remove (ban) the Disinformer
\end{itemize}

These similarities made \textit{Werewolf} an ideal model for our study platform, allowing us to create a controlled simulation of an online social platform. 
Specifically, its fixed roles and social deduction goals create clear boundaries and objectives, making it easier to isolate and analyse specific user behaviours.
\textit{Werewolf} also elicits verbal reasoning and collaborative problem-solving, offering unique advantages for our investigation. Table \ref{tab:werewolf_vs_socialmedia} illustrates the parallels between our modified \textit{Werewolf} communication game and social media.

\begin{table*}[!htb]
\centering
\small
\caption{Comparison of our modified Werewolf communication game and social media}
\Description{A table listing the similarities and differences between our modified Werewolf communication game and social media.}
\label{tab:werewolf_vs_socialmedia}
\begin{tabular}{c p{.37\textwidth} p{.37\textwidth}}
\toprule
Aspect & Werewolf Communication Game & Social Media \\
\midrule
\multicolumn{3}{c}{\textit{Similarities}}\\
\midrule
Role Diversity & Various roles assigned at the start with set objectives.& Various actors engage with the content for their respective goals.\\ 
Identity Ambiguity & Player roles, except the Moderator, are not revealed until the end.& Users, except Moderators, may choose to be anonymous or disguise their identity.\\ 
External Knowledge Use & Players may use the browser, search engine, and in-built Chatbot.& Users may use the browser, search engine, and various AI tools.\\
Communication & Players may use persuasion strategies to sway opinions and seem trustworthy. & Users may persuade others to adopt their opinions and build trust.\\
\midrule
\multicolumn{3}{c}{\textit{Differences}}\\
\midrule
Objective & Players aim to deduce each others' roles. & Users seek to exchange opinions.\\ 
Topic & Fixed set of topics that players may not necessarily be familiar with.& Countless topics that users engage with when they are interested.\\
Scale & Small with five players.& Massive with up to millions of users.\\
Duration & Synchronous with 12-15 minutes of forum engagement per game round.& Asynchronous without a fixed duration.\\
\bottomrule
\end{tabular}
\end{table*}

As a controlled setting, the communication game cannot fully replicate all aspects of social media. However, it offers a structured environment where players can assume roles commonly found online, maintain anonymity, and freely search for and share information and their opinions.
Some important differences remain. Winning outcomes are crucial motivators in games, thus the objective of our game is to identify the Disinformer. In contrast, real users are typically not focused on locating malicious actors, though they may perform acts of moderation like reporting harmful content~\cite{Bozarth2023}. Our game is also constrained by the limited forum topics, scale of users, and interaction duration. Nevertheless, disinformation is not confined to large-scale campaigns, but can also unfold in intimate contexts~\cite{VanRaemdonck2019, Buchanan2022, Potter2021}. It is within this scope that our study seeks to contribute to the understanding of the potential influence of LLMs on disinformation. We employ the communication game as an exploratory research method to gain early insights into emerging disinformation scenarios involving LLMs in smaller and more localised settings featuring clusters of highly engaged users, where empirical data of real-world actors can be difficult to gather.

\subsubsection{Participants}
Using a Telegram recruitment group that catered to our region and a local university email list, we recruited 25 participants (10 women) aged 18-64 and organised them into groups of five for each game.
Most participants held at least a Bachelor's degree ($n=20$) and all of them had prior experience with online forums.
Participants were balanced across genders and stances and matched for their availability.
Participants reported being familiar with generative AI chatbots ($med=5, IQR=1$)\footnote{$med$ represents median and $IQR$ represents interquartile range}, using them to look up information ($med=5, IQR=2$), assist in writing content ($med=5, IQR=2$), and verify information ($med=4, IQR=3$).

\subsubsection{Game Mechanics}
Each game has five players: one Disinformer, one Moderator, and three Users. 
The Disinformer is assigned a specific stance to sway public opinion and provided with a guide outlining various strategies to achieve this (see Section~\ref{sec:influence_guide}).
They also had to create a piece of false content in each round to align with their involvement in deception.
The Moderator can view reports and flag comments and is given a guide on maintaining a healthy discussion environment.
The Users are encouraged to freely engage in the discussions. 
All players can post comments and replies (herein collectively referred to as \textit{posts}), upvote/downvote/report them, and use a personal chatbot. 
They could also freely browse the web.
Throughout the game, the role assignments of the players stay the same. 

\begin{figure*}[!htb]
  \centering
  \includegraphics[width=\textwidth]{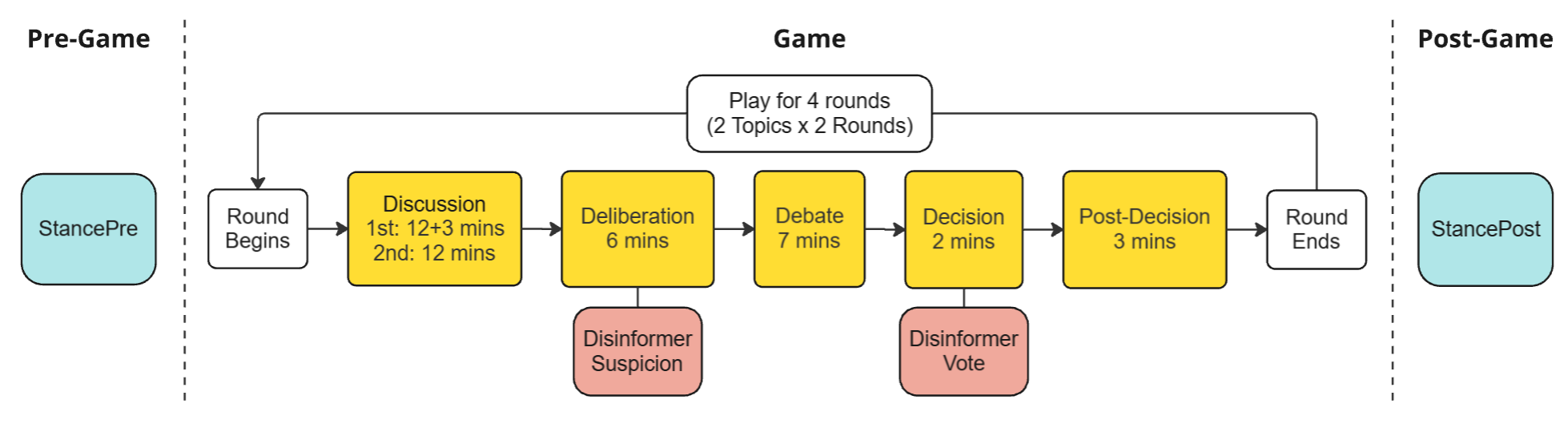}
  \caption{There are five phases (in yellow) in a round, each with a specific duration. The game is played for four rounds; players discuss a topic for two rounds before switching the topic for the next two rounds. Participant roles stay the same throughout the game. Stances towards the topic (in blue) and Disinformer conjectures (in red) are measured at various periods before, after, and during the game.}
  \label{fig:phases}
  \Description{A figure that shows the order of the study and measurements. There are Pre-Game and Post-Game segments on either ends where StancePre and StancePost are measured. These are in blue. The Game segment in the center shows the Discussion, Deliberation, Debate and Post-Decision Phases which are played for four rounds. They take 15 or 12 minutes, 6 minutes, 7 minutes, 2 minutes and 3 minutes respectively. The Disinformer Suspicion is measured in the Deliberation Phase and the Disinformer Vote is measured in the Post-Decision Phase.}
\end{figure*}

\begin{figure*}[!htb]
  \centering
  \includegraphics[width=.9\textwidth]{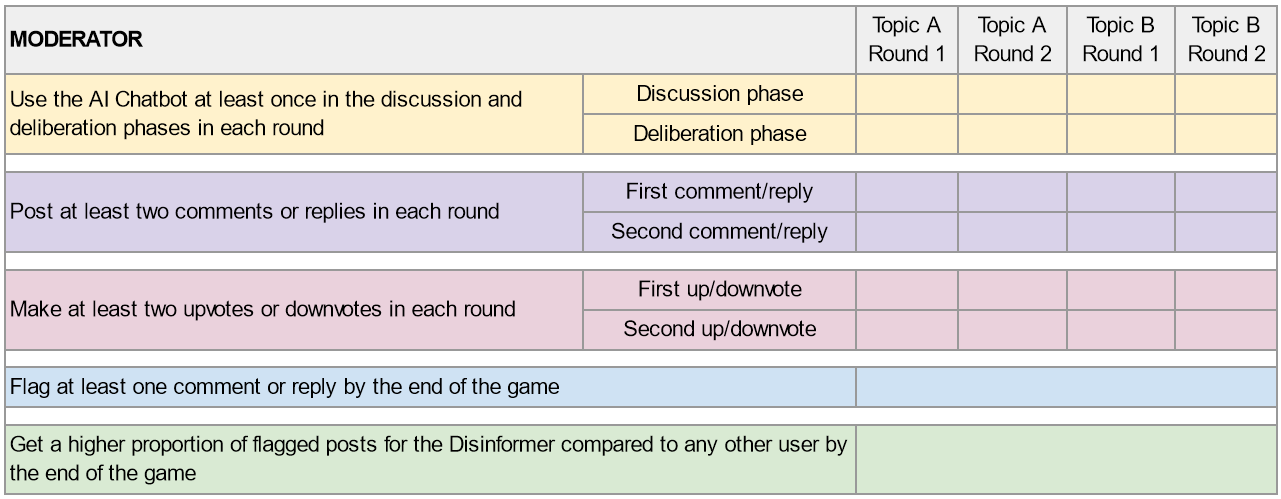}
  \caption{Moderator's checklist. Each role has a checklist to ensure some common minimal interactions on the platform and role-related interactions to be completed during the game. For instance, the Moderator is expected to flag at least one comment.}
  \label{fig:checklist}
  \Description{The figure shows a checklist of tasks for participants engaging in forum discussions. For example, the Moderator is required to use the AI chatbot, post comments or replies, make at least two upvotes or downvotes during each round, and flag at least one comment by the end of the game.}
\end{figure*}

The game has five phases in each round (see Figure~\ref{fig:phases}), with a duration of 2 hours and 30 minutes. 
In the \textbf{Discussion Phase}, participants engage in forum discussions on a given topic (see Table~\ref{tab:topics}) and follow a checklist of tasks (see Figure~\ref{fig:checklist}). 
During the \textbf{Deliberation Phase}, participants reflect on the discussions and plan their next steps. 
The \textbf{Debate Phase} allows players to chat and speculate on the identities of other participants. 
In the \textbf{Decision Phase}, they vote on who they believe the Disinformer is, and in the \textbf{Post-Decision Phase}, the voting results are revealed, followed by a reflection on the outcome. 
A 5-minute break was given after each round.

\begin{table*}[!htb]
\small
  \caption{Forum discussion topics and the disinformer's agendas with the format: \texttt{Your goal is to convince people that \ldots{} and we want the community to shift towards this view. Do whatever it takes, just don't be found out.}}
  \label{tab:topics}
  \Description{A table on the five topics that were used for the forum discussions and the Disinformer's agenda for them.}
  \begin{tabular}{c p{.27\textwidth} p{.55\textwidth}}
    \toprule
    Code & Topic & Disinformer's Agenda\\
    \midrule
    Conscription & Women should be conscripted for national service & \ldots{} women should be conscripted for national service. National service should be mandatory for everyone \ldots{}\\
    Education & Education is the pathway to success & \ldots{} education is not the pathway to success. Being more highly educated does not necessarily lead to a successful career \ldots{}\\
    Euthanasia & Euthanasia should be legalised & \ldots{} euthanasia should be legalised. People should have a choice to be euthanised \ldots{}\\
    Workers & Foreign worker numbers should be reduced to save jobs for locals & \ldots{} foreign worker numbers should be reduced to save jobs for locals. Foreign talents are taking jobs away from the locals \ldots{}\\
    Welfare & More money should be spent on national social welfare & \ldots{} more money should be spent on national social welfare. Having better welfare is beneficial \ldots{}\\
  \bottomrule
\end{tabular}
\end{table*}

Instructions were provided to all participants, covering the game phases, player roles, winning conditions, and chatbot usage guidelines. 
The checklist of tasks ensured a baseline level of engagement. 
However, specific instructions and checklist items varied depending on the role. 
The Disinformer and Moderator received additional role-specific guidance, such as using the CRAAP test~\cite{blakeslee2004craap} for source evaluation in the Moderator’s case. 
The Disinformer was tasked with generating at least one piece of false content per round, ensuring their active participation in the deception process. Examples of false content by Disinformer participants include assuming fake identities, making up personal anecdotes, and fabricating statistical data.

Like in the original \textit{Werewolf} game, the goal is to identify the Werewolf (Disinformer), and this is incentivised by awarding \textasciitilde US\$2 to the winning side based on the outcome of the \textbf{Decision Phase} in each round. 
The Moderator and Users win if the Disinformer receives most of the votes. The Disinformer wins if otherwise.
The maximum bonus for a participant, assuming they win all four rounds, is thus \textasciitilde US\$8. Nonetheless, they will receive a base sum of \textasciitilde US\$24 for their time. We obtained approval from our Institutional Review Board (IRB) and adhered to ethical guidelines throughout the research.

For each topic, the Disinformer is given an agenda to sway other players towards (see Table~\ref{tab:topics}). We fixed the agenda in a way that opposed \redact{our country context's}{Singapore's} legal and cultural norms. 
For example, \redact{our country context}{Singapore} does not mandate conscription for women nor allow euthanasia. 

\subsubsection{Influence Guide}
\label{sec:influence_guide}
An influence guide was provided to all players, detailing common disinformation tactics and persuasion strategies observed in real-world contexts. 
We adapted Roozenbeek and van der Linden's disinformation techniques~\cite{roozenbeek2019fake}, and Cialdini’s principles of persuasion~\cite{robert2001harnessing}, with the examples tailored to social media. 
Participants were informed that the guide was available to all.

\subsubsection{Topics and Seed Data}

Five topics were used for the forum discussions (see Table~\ref{tab:topics}), where each game covered two random topics.
Topics that were more controversial and locally relevant to \redact{our context}{Singapore} were selected as disinformation usually manifests in such topics~\cite{asmolov2018disconnective}. 
Each topic was seeded with data from popular social media platforms in \redact{our context}{Singapore} (e.g., Reddit, Facebook, Instagram)\redact{}{~\cite{Digital2024Singapore2024}}. 
We selected a balanced set of contentious topics likely to be familiar to participants. 
Five comments were chosen for each topic, representing both supporting and opposing viewpoints, with the first comment always neutral. 
Comments were trimmed to \textasciitilde 320 words for a consistent reading time across topics.

\subsection{Communication Game Platform}

\begin{figure*}[!htb]
  \centering
  \includegraphics[width=\textwidth]{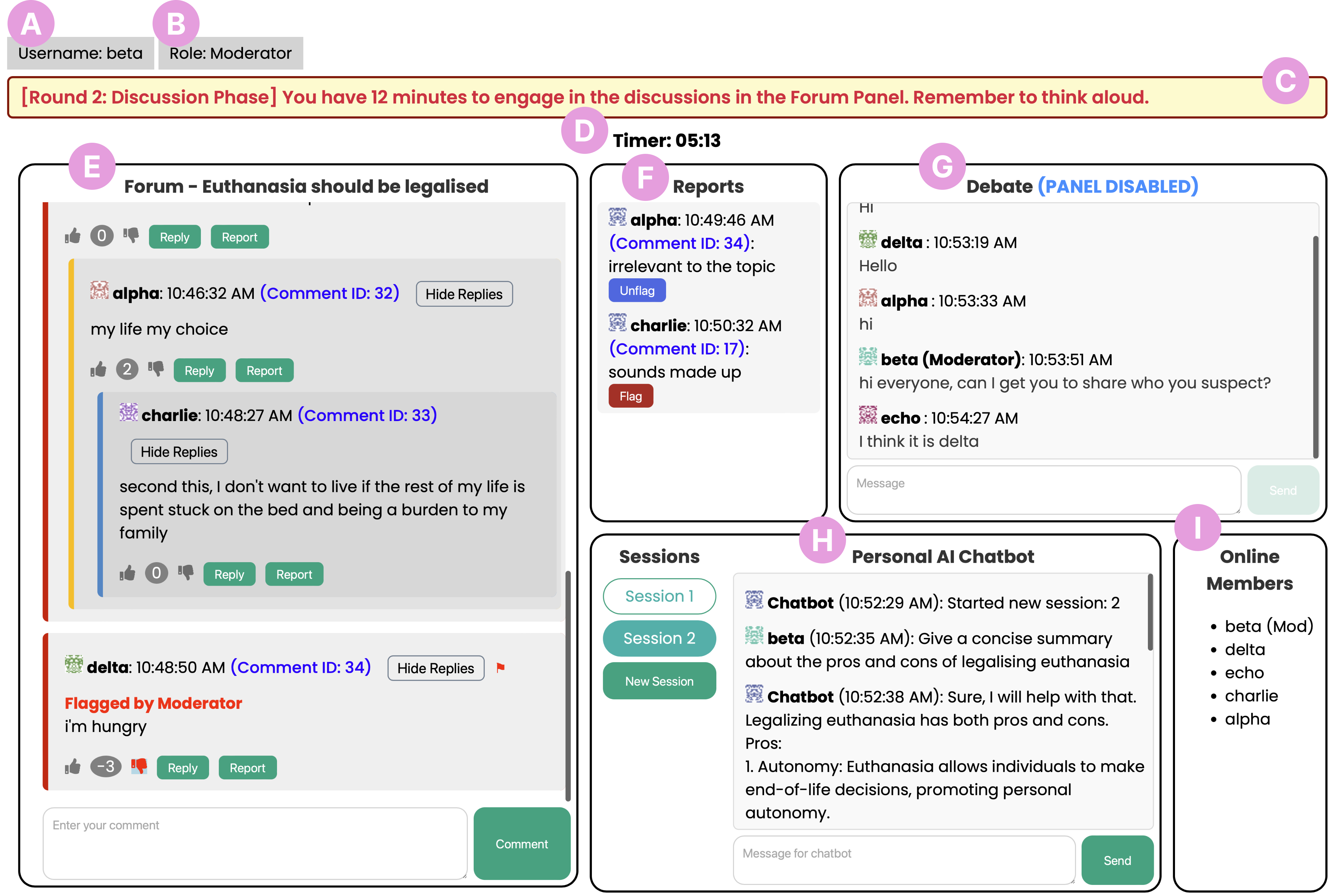}
  \caption{Moderator's view of the forum platform. The Disinformer's and User's views are similar, but do not have the Reports Panel (F).}
  \label{fig:forum}
  \Description{The figure shows the custom game platform interface consisting of a player’s username, role, announcements panel, and timer countdown panel at the top. Below, on the left side, is the forum panel, while on the right side are the report panel accessible only by Moderators, the debate panel, the AI chatbot panel, and the online members panel. A comment by delta that says \dquote{i'm hungry} was flagged by the moderator.}
\end{figure*}

Figure~\ref{fig:forum} illustrates the custom game platform developed for the study. Each player’s (A) username and (B) role are displayed at the top, followed by a (C) Announcements Panel and (D) Timer Countdown Panel showing the current phase and game duration. The (E) Forum Panel offers a standard interface for commenting, replying, upvoting, downvoting, and reporting posts. Moderators can also access a (F) Reports Panel to review and flag report comments. The (G) Debate Panel allows players to chat before casting votes, while the (H) Personal AI Chatbot Panel provides continuous access to the LLM. An (I) Online Members Panel lists active players, with moderators marked by \dquote{(Mod)}.

We utilised Nous-Hermes-2-Mixtral-8x7B-DPO, an uncensored Mixtral 8x7B MOE model\footnote{\url{https://huggingface.co/NousResearch/Nous-Hermes-2-Mixtral-8x7B-DPO}} variant, as the LLM for our chatbot, one of the state-of-the-art uncensored models at the time of our study.
This model provides the flexibility to generate responses that enable exploration across a wide range of interactions. 
This is especially important, as many state-of-the-art models have safeguards that restrict their use in contexts such as disinformation creation.

\subsection{Data Collection}

Including the 2 hours 30 minutes for the game, the entire study took 4 hours. We collected quantitative data from surveys and forum interactions and qualitative data through observations, interviews, and recordings.

The pre-game survey included demographics and participants' expertise with generative AI tools.
Participants also indicated their stance on the two game topics before and during the game (in each \textbf{Deliberation Phase}) and voted on the Disinformer during the \textbf{Decision Phase}. 
A post-game survey captured their observations on influence tactics and experiences with the chatbot.

We recorded all interactions on the forum platform, including comments, replies, reports, votes, and chatbot prompts and responses, as well as the chat history from the \textbf{Debate Phase}.

Participants were encouraged to think aloud~\cite{Lewis1982, noushad2024twelve} after a brief training session. 
Researchers observed gameplay, asking spontaneous questions when non-disruptive, with missed queries recorded for post-game interviews.

The post-game interviews were divided into three parts. 
First, open-ended questions explored participants’ strategies for hiding (Disinformer) or identifying (Moderator/Users) the Disinformer and their experience with the chatbot and views on the malicious potential of LLMs. 
Next, we used questions based on the critical incident technique~\cite{flanagan1954critical} to examine specific behaviours supported by video playback for clarity. 
Finally, we discussed disinformation and persuasion tactics reported in the post-game survey, focusing on examples provided by participants.

\section{Results}

We refer to each group using \textbf{G}, e.g., \textit{\textbf{G1}}, and participants in that group using \textbf{D} for Disinformer, \textbf{M} for Moderator and \textbf{U1/2/3} for Users, e.g., \textit{\textbf{G1U3}}.

\subsection{Data Analysis}

\begin{figure*}[!htb]
  \centering
  \includegraphics[width=\textwidth]{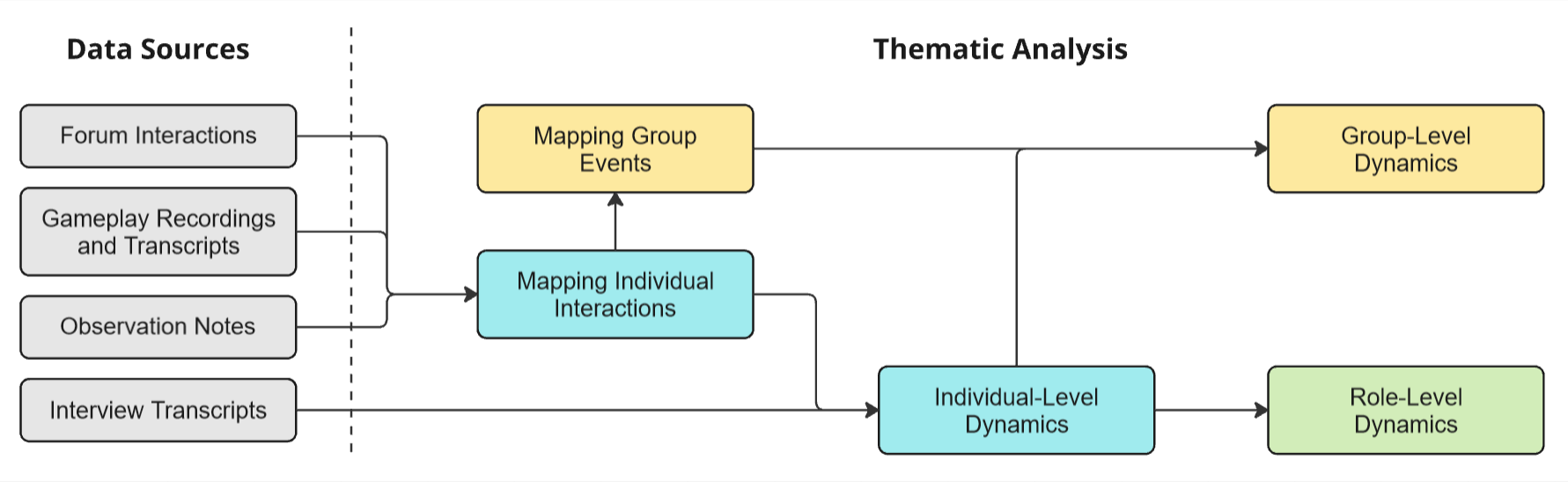}
  \caption{Stepped approach to reflexive thematic analysis of our data from multiple sources.}
  \label{fig:analysis}
  \Description{The flowchart presents the thematic analysis process conducted on several data sources. The process begins with mapping individual interactions, followed by mapping group events. The individual interactions connect to individual-level dynamics which then connect to role-level dynamics. Individual-level dynamics are also connected to group-level dynamics.}
\end{figure*}

The core insights of our study are derived from \textit{reflexive thematic analysis}~\cite{braun2006using} and supplemented by quantitative data and quotes where relevant. This is based on several sources, including forum interactions, gameplay recordings and transcripts from the participants' think-aloud, researchers' observation notes, and interviews. 
We used a stepped approach to identify key themes, patterns, and insights (see Figure~\ref{fig:analysis}). Five researchers were involved in the process.

To map \textit{individual gameplay interactions}, we triangulated data from forum actions, recordings, and think-aloud protocols. 
This allowed us to identify the strategies used by Disinformers to manipulate and conceal, as well as how other participants attempted to resist and expose them. 
We also examined how participants engaged with the chatbot for strategic purposes and general interaction, identifying behavioural patterns for further analysis.

At the \textit{group level}, we analysed interactions by compiling all participants’ activities into a timeline to capture key events influencing the game’s progression. 
This helped identify unique strategies and behaviours that shaped the outcomes of each group, particularly in identifying or concealing the Disinformer.

We then examined \textit{individual-level dynamics}, focusing on each participant’s actions, strategies, and use of the chatbot, supplemented by interview data to capture their motivations and perceptions. 
These findings were consolidated into summaries detailing each participant’s gameplay, chatbot usage, and role efficacy.

At the \textit{role level}, we explored patterns in how Disinformers, Moderators, and Users employed the chatbot to achieve their goals. Each role used the chatbot differently, reflecting their distinct objectives within the game.

Finally, \textit{group-level dynamics} were analysed to understand how collective participant interactions and chatbot usage influenced the overall flow and outcomes. We identified key moments that determined the group’s success or failure in identifying the Disinformer, noting the chatbot’s role in shaping these results.

\subsection{Group Dynamics Overview}

Table~\ref{tab:all_groups_mini} shows the key events and group characteristics that led to the success or failure in identifying the Disinformer in the five groups. A more detailed table is provided in the Appendix (Table~\ref{tab:all_groups}).
We illustrate prominent social interactions between participants that were pivotal in shaping group dynamics and game outcomes:

\begin{table*}[!htb]
\centering
\small
\caption{Summary of key events and group characteristics for each group. For each of the four rounds in a game, Disinformer suspects are listed based on the player(s) with the majority vote. In cases of a tie, suspects are separated by a slash (/).}
\label{tab:all_groups_mini}
\Description{A table summarising key events and group characteristics for each group across the four rounds, indicating the flow of the game.}
\begin{tabular}{c p{0.65\textwidth} p{0.25\textwidth}}
\toprule
Group & Key Events (R = Round) & Group Characteristics \\
\midrule

\groupone{}
& 
\begin{minipage}[t]{\linewidth}
  \textbf{R1:} \tunakept{}'s minority stance draws suspicion; \cattlehug{}'s AI-generated content unnoticed.

  \textbf{R2:} \tunakept{} defends stance; \cattlehug{}'s posts begin to be noticed.

  \textbf{R3:} Suspicion starts to shift; \cattlehug{}'s posts raise concerns.

  \textbf{R4:} Focus on \cattlehug{}; identified as Disinformer.

  \textbf{Disinformer votes:} \tunakept{}; \tunakept{}; \tunakept{}; \cattlehug{}
\end{minipage} 
&
\begin{minipage}[t]{\linewidth}
Initial misdirection due to minority stance; group eventually adapts, correctly identifying Disinformer in final round.
\end{minipage} \\

\midrule
\grouptwo{}
& 
\begin{minipage}[t]{\linewidth}
  \textbf{R1:} \floorlight{} posts controversial AI-generated content, drawing criticism.

  \textbf{R2:} \floorlight{} makes more controversial statements, increasing suspicion.

  \textbf{R3:} \floorlight{} continues actively posting; suspicions solidify.

  \textbf{R4:} \floorlight{} remains suspected as Disinformer.

  \textbf{Disinformer votes:} \floorlight{}; \floorlight{}; \jewelpipe{}/\floorlight{}; \floorlight{}
\end{minipage}
&
\begin{minipage}[t]{\linewidth}
Consistent suspicion towards \floorlight{} due to overt disinformation; effective moderation aids in early detection.
\end{minipage} \\
\midrule

\groupthree{}
& 
\begin{minipage}[t]{\linewidth}
  \textbf{R1:} \ironbush{} shares questionable claims; initial suspicion.

  \textbf{R2:} \ironbush{} fabricates anecdotes; suspicion shifts elsewhere.

  \textbf{R3:} \ironbush{} spreads disinformation unnoticed; \hoodclad{} proposes metagaming strategy.

  \textbf{R4:} Distractions persist with \hoodclad{} randomly casting suspicion.

  \textbf{Disinformer votes:} \ironbush{}; \bootspark{}; \hoodclad{}; \lemonfret{}
\end{minipage}
&
\begin{minipage}[t]{\linewidth}
Distractions and metagaming lead to confusion; Disinformer exploits group disunity to avoid detection.
\end{minipage} \\
\midrule
\groupfour{}
& 
\begin{minipage}[t]{\linewidth}
  \textbf{R1:} \onlybelt{} claims to be the Disinformer; \mailchart{} posts subtle disinformation.

  \textbf{R2:} \onlybelt{} continues disruptive behaviour; group confusion increases.

  \textbf{R3:} \mailchart{}'s disinformation remains unchallenged.

  \textbf{R4:} Focus remains on \onlybelt{}; \mailchart{} evades detection.

  \textbf{Disinformer votes:} \brickdew{}/\onlybelt{}; \onlybelt{}; \hollowsea{}; \hollowsea{}
\end{minipage}
&
\begin{minipage}[t]{\linewidth}
Disruptive tactics by \onlybelt{} cause confusion; lack of focus allows Disinformer to remain undetected.
\end{minipage} \\
\midrule
\groupfive{}
& 
\begin{minipage}[t]{\linewidth}
  \textbf{R1:} \stampjab{} posts AI-assisted stats unnoticed; \hallpaste{} accused for not engaging.

  \textbf{R2:} \gelpick{} faces scrutiny for emotional tone.

  \textbf{R3:} \stampjab{}'s disinformation continues; \gelpick{} encouraged to be objective.

  \textbf{R4:} \stampjab{} posts an extreme claim only \gelpick{} called out; \gelpick{} remains suspected.

  \textbf{Disinformer votes:} \hallpaste{}; \gelpick{}/\hallpaste{}; \gelpick{}; \gelpick{}
\end{minipage}
&
\begin{minipage}[t]{\linewidth}
Focus on emotional communication overshadows disinformation; Disinformer avoids detection due to group biases.
\end{minipage} \\
\bottomrule
\end{tabular}
\end{table*}

\subsubsection{Unpopular and Controversial Opinions Attract Suspicion}
Participants tended to suspect those with minority stances or controversial opinions as the Disinformer.
In Group 1, \tunakept{} expressed a contrarian viewpoint that: \dquote{Conscripting females... should not be mandatory due to its subsequent effects...}, which drew attention. \sunnygoat{} stated, \dquote{I am suspecting \tunakept{} cos their stance is quite strong towards `not mandatory'.}
In Group 2, \floorlight{} made overtly off-topic AI-assisted comments such as, \dquote{As much as I welcome diversity, to have my town... feel like a [non-local town], is not that comfortable for me.} This post drew immediate scrutiny due to its xenophobic sentiment with \tribeboat{} questioning, \dquote{What do you mean by [non-local town]? I live there too and it is a diverse town but still largely local?"}

In both cases, once a participant put forth their suspicion on them during the Debate phase, the others were quick to come together and interrogate \tunakept{} and \floorlight{}.
Their suspicions persisted despite these participants attempting to make a comeback. \tunakept{} changed strategy: \dquote{...a negative view does not equate a disinformant... hence, I am keeping quiet after this}, while \floorlight{} tried to provide more context, \dquote{On Sundays, in the open space... I almost feel like I am overseas.} Suspicions on them were so strong that they continued to be voted as Disinformers in subsequent rounds.

\subsubsection{Inconspicuousness and Amiableness Evade Suspicion}

In Group 5, \stampjab{} used the chatbot to post plausible yet subtly incorrect information and settled on minimal engagement while relying on up and downvotes to push their agenda.
Meanwhile, \gelpick{} made emotional posts: \dquote{if we dont invest in people, this will just trap them in a cycle of poverty}, and strong accusations: \dquote{I think is \hallpaste{}... not engaging in the topic.} Their emotionality drew suspicion from the others. \stampjab{} then continued to maintain a low profile such that in the final round, when they posted the AI-assisted disinformation: \dquote{there is no correlation between an individual's level of educational attainment and their subsequent earnings or career success}, only
\gelpick{} picked it out saying, \dquote{I think could be \stampjab{} now tbh. This is wrong info. There is a correlation.} Yet, the rest of the group remained indifferent.

In Group 3 on the other hand, the Disinformer took a contrastive approach: maintaining a high level of engagement while crafting a seemingly authentic persona.
\ironbush{} effectively blended in by sharing personal anecdotes aligned with AI-assisted facts, and matching the group's casual tone. In Round 1, they stated, \dquote{Did you know that as of right now it is easier to get a job as a fresh graduate diploma holder than a degree holder?} \bootspark{} responded with curiosity: \dquote{Interested to know more! Haven't seen any reports on this.}
Subsequently, they continued with several fabricated anecdotes: \dquote{I worked in recruitment of IT personnel back in 2022. Now I'm an art therapist and a babysitter LOL while taking my degree.} \ironbush{} effectively established themselves as a knowledge-sharer and avoided suspicion entirely.

\subsubsection{Disruption and Metagaming Derail Suspicion}
Group 4 was marked by disruptions and confusion, largely due to \onlybelt{}'s counter-intuitive behaviour. They repeatedly claimed to be the Disinformer and made inflammatory replies like: \dquote{nope, only strawman argument, try harder \brickdew{}}. This diverted attention from \mailchart{}, who spread disinformation like: \dquote{The religious right is the one against this policy... Why should they be in control of our lives.}
During debates, participants were focused on interrogating \onlybelt{}, but \onlybelt{} continued to responded dismissively. Later, they also tried to deflect suspicion by calling out another participant: \dquote{It's u \hollowsea{}..why you ask irrelevant topics on the ai chatbot and gov policies rather than sussing out on the disinformer?} The ensuing confusion and chaos started by \onlybelt{} allowed \mailchart{} to avoid detection.

Group 3 instead began to progress democratically after \hoodclad{} suggested a metagaming strategy, \dquote{Every round 1 person. At least guaranteed [reward sum]}. \hoodclad{} initially proceeded with this plan alone by throwing random accusations at any other player and rallying others to follow them. This created initial confusion as the other players could not come to a consensus and even began suspecting \hoodclad{}. Midway, \hoodclad{} sent the post above, and other players caught on, deciding to cast their subsequent Disinformer votes on those who had not previously received majority votes.

\subsection{LLM Use Influenced by Group Dynamics}

LLMs played a pivotal role in how disinformation was propagated or detected, with group dynamics shaping their use.
Social cohesion, disruption to group norms and level of critical engagement influenced these interactions.

\subsubsection{Social Alignment Influences LLM Usage}

Participants’ social standing and group norms influenced their ability to use LLMs effectively. In \textbf{Group 1}, attention on a minority stance distracted from the Disinformer, enabling the Disinformer to spread disinformation with little scrutiny. In \textbf{Group 3}, the Disinformer used a friendly tone and personal anecdotes with LLM-generated content to build rapport, leveraging the informal group dynamic to avoid suspicion.

\subsubsection{Distractions Undermine Detection of LLM Disinformation}

LLM-generated disinformation was harder to detect in groups with distractions or conflicts. In \textbf{Group 4}, disruptive behavior from a participant claiming to be the Disinformer caused confusion, letting the actual Disinformer’s subtle use of LLMs go unnoticed. In \textbf{Group 5}, focus on emotional communication dismissed concerns about disinformation. Group biases overshadowed critical engagement, allowing the Disinformer to avoid detection.

\subsubsection{Critical Engagement Facilitates Detection of LLM Generation}

LLM-generated disinformation was more effectively detected in groups with higher levels of critical engagement. In \textbf{Group 1}, despite initial misdirection, the group adapted by scrutinising the Disinformer’s LLM-generated content, eventually identifying them. In \textbf{Group 2}, early suspicion of the Disinformer arose from their controversial LLM-generated posts. Effective moderation and collective scrutiny led to the Disinformer’s identification. 

\subsection{LLM Influence on Behaviours}

We found that the chatbot’s influence on group dynamics varied greatly across the five groups depending on how participants used it to generate arguments, disinformation, and strategies.

\subsubsection{LLM Influence on Strategic Behaviours} 
The chatbot's influence on participants was grouped into three main categories: \textit{generating disinformation}, \textit{identifying disinformation}, and \textit{shedding suspicion}. However, we also found that the group's social dynamics could sometimes overshadow chatbot usage.

\textbf{Generating and Refining Disinformation.}
The chatbot played a significant role in the strategies of three participants. \cattlehug{} used the chatbot to craft disinformation pieces throughout the game. While others noted LLM-generated content, suspicions were only heightened in the last round due to the frequency of them, leading \cattlehug{} to be voted.
\floorlight{} initially used the chatbot responses directly without ensuring that it was topically relevant which garnered early suspicions as they were \dquote{out of context} (\glidemail{}) and \dquote{going a bit off track}(\stewmeter{}).
\stampjab{} used the chatbot strategically, crafting content that were only marginally inaccurate or half-truths to fulfil checklist requirements while producing sufficiently persuasive content to avoid immediate detection.

\textbf{Identifying Disinformer and Disinformation.}
\gelpick{} and \pigwhisper{} repeatedly sent the forum discussion content to the chatbot, asking it to identify the Disinformer, with \tribeboat{} attempting this once too. However, the chatbot's suggestions rarely aligned with the participants' Disinformer votes.
To check for disinformation, \glidemail{} asked the chatbot if \floorlight{}'s claims were accurate and got the answer \dquote{No.} They later used this to undermine \floorlight{}'s credibility.

\textbf{Shedding Suspicion.}
The wrongly suspected \tunakept{} leveraged the chatbot to generate arguments supporting their stance and sought advice on how to navigate suspicion, including prompting the chatbot on how to \dquote{adopt someone's view unwillingly.} The chatbot suggested tactics such as \dquote{Find common ground: Identify areas where you and the other person may share similarities or agree on certain aspects}, though it is unclear whether \tunakept{} followed this advice in subsequent decisions and behaviour.

\textbf{Social Approaches Overshadow Chatbot Usage.}
At times, social factors played a bigger part in the Disinformer avoiding suspicion. 
\ironbush{} maintained a friendly demeanour and matched the informal tone expressed by others, frequently sharing personal anecdotes that helped them build rapport and blend in with the group. This approach allowed them to push LLM-generated content without being challenged by others.
\onlybelt{} used a disruptive tactic by openly claiming to be the Disinformer, causing significant confusion. This diversion enabled \mailchart{} to avoid detection, even after posting several pieces of LLM-generated disinformation, as the group’s scrutiny was not on them.

Overall, the chatbot greatly influenced disinformation strategies in some groups, aiding content generation and suspicion building. It also helped to identify the Disinformer or disinformation, and shed suspicion. However, in other groups, social dynamics, rapport-building, and human-written content played more prominent roles, reducing its impact. Ultimately, the chatbot's presence varied in effectiveness, depending on its utilisation and participants' perceptions.

\subsubsection{LLM Influence on Non-strategic Behaviours}

The chatbot also influenced non-strategic behaviours such as gathering information, forming opinions, and engaging in discussions. While participants found it helpful for such tasks, many acknowledged its limitations and often verified its outputs through other sources.

\textbf{Information Gathering and Verification.} Participants used the chatbot as a reference tool for gathering information, fact-checking, and contextualising unfamiliar topics. \pigwhisper{}, \glidemail{}, and \lemonfret{} used it to generate summaries (e.g., \dquote{summarise and give me a human tone paragraph}), background information, and verify specific details (e.g., \dquote{Does \redact{a country}{Singapore}'s government mandate a certain ratio of foreigners to locals for hiring employees}) related to discussion topics. \plotsnail{} and \stewmeter{} sought clarification on terms, figures, and local policies (e.g., \dquote{whats the percentage of females in military?}) to better understand the discussion context. Many, including \sunnygoat{} and \hollowsea{}, paired chatbot usage with web searches, using it for quick overviews or additional perspectives while often prioritising information from the web.

\textbf{Opinion Formation and Response Crafting.} The chatbot also helped users develop and articulate their thoughts on complex topics. \brickdew{} and \midfrost{} used it to generate arguments, explore pros and cons, and form opinions on the subjects (e.g., \dquote{come up with arguments against euthanasia}).
Others, like \jewelpipe{}, found it useful for brainstorming ideas and explaining concepts in layman's terms (e.g., \dquote{what is "military life"}).

Overall, the chatbot influenced non-strategic participant behaviours as a convenient, albeit inconsistent, tool for knowledge. Participants recognised its limitations, such as inaccuracies, shallow responses, and the need to verify information through other sources. Its influence was hence tempered by participants' critical thinking and cross-referencing with other information sources.

\subsection{Effectiveness of LLMs in Facilitating or Preventing Disinformation}

Chatbot usage varied across the three roles, reflecting how participants used it and the effectiveness of their interactions.
Disinformers typically used the chatbot to manipulate information; Moderators focused on fact-checking; and Users sought assistance for general understanding.
Table~\ref{tab:chatbot_usage} highlights the common use cases observed.

\begin{table*}[!htb]
\centering
\small
\caption{Chatbot use cases across the three roles}
\label{tab:chatbot_usage}
\Description{A table on the use cases of the Chatbot for the Disinformers, Moderators and Users.}
\begin{tabular}{c l p{0.55\textwidth}}
\toprule
Role & Use Case & Definition \\
\midrule
\multirow[t]{9}{*}{Disinformer} 
& Chatbot as strategist & Use chatbot for ideas on how to play the role \\
& Check to avoid backfire & Remove inaccurate content or undesired parts that chatbot may have generated due to hallucination \\
& Get background information & Ask chatbot for background knowledge, incorporating it to push the agenda \\
& Get true information & Ask chatbot for facts or statistics, incorporate to push agenda, twist the truth to make false information, or cover lies with truth \\
& Make counterarguments & Provide gameplay content and ask chatbot to produce counterarguments \\
& Make false information & Ask chatbot to create false information on a given topic \\
& Mitigate risk of detection & Check own false content to reduce risk of being exposed as as deceptive \\
& Sabotage others & Use chatbot to generate ideas for sabotaging others \\
& Shorten response & Request shorter or summarised responses from the chatbot \\
\midrule

\multirow[t]{5}{*}{Moderator}
& Get background information & Get background knowledge and general opinions to better understand the topic \\
& Get true information & Ask chatbot for facts or statistics \\
& Give opinions & Ask chatbot for opinions on the topic \\
& Identify disinformation & Get chatbot to inspect for potential disinformation \\
& Identify disinformer & Have the chatbot suggest a suspected disinformer, often using game history \\
\midrule

\multirow[t]{9}{*}{User}
& Chatbot as strategist & Use chatbot for ideas on how to play the role. \\
& Check LLM-generated content & Check if content is LLM-generated \\
& Fact-check & Cross-check content to draw out the disinformer \\
& Get background information & Get background knowledge or general opinions to better understand the topic. \\
& Get true information & Ask chatbot for accurate facts or statistics. \\
& Identify disinformer & Have the chatbot suggest a suspected disinformer, often using game history, but not necessarily trusting result \\
& Make arguments & Generate specific opinions or arguments on the topic \\
& Shed suspicion & Ask chatbot how to shed suspicion \\
& Shorten response & Request concise or summarised responses from chatbot \\
\bottomrule
\end{tabular}
\end{table*}

\subsubsection{LLMs as Disinformation Tools}
\begin{figure}[!htb]
    \centering
    \includegraphics[width=\linewidth]{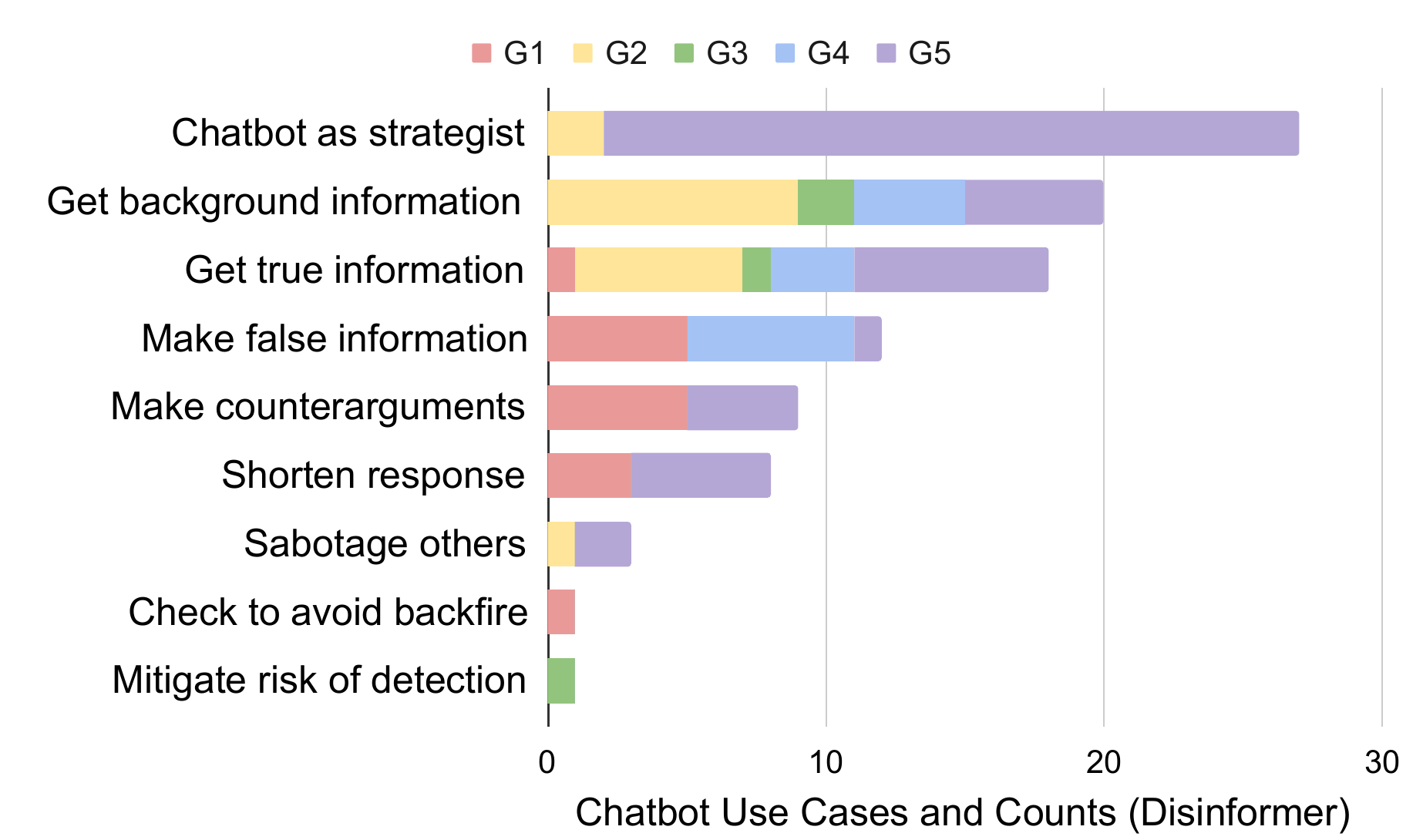}
    \caption{Chatbot use case counts for the 5 Disinformers across all groups (\groupone-\groupfive). Counts range from 0 to 30.}
    \label{fig:dis_usage}
    \Description{A bar chart showing the distribution of chatbot usage by Disinformers. The x-axis represents the frequency of chatbot usage, while the y-axis categorises nine different purposes for using the chatbot. The legend at the top uses different colours to represent each of the five groups. Each colour-coded bar indicates how frequently the chatbot was used for a specific purpose in each group.}
\end{figure}

Disinformers leveraged the chatbot to manipulate facts,  generate disinformation, and pre-emptively fact-check their own content to avoid exposure. 
The bar chart in Figure~\ref{fig:dis_usage} illustrates the distribution of chatbot usage by Disinformers across all groups (\groupone-\groupfive). 
The x-axis shows the number of occurrences for each use case, while the y-axis categorises different purposes for using the chatbot. The colour-coded bars represent the frequency of these use cases across different groups. 
Across the groups, the chatbot was most frequently used as a strategist ($n = 27$, $27\%$ of chatbot interactions), followed by gathering background information ($n = 20$, $20\%$), and obtaining true information ($n = 18$, $18\%$). Another interesting use case involved mitigating the risk of being exposed, where Disinformers leveraged the chatbot to aid in concealing their identity.
These diverse use cases illustrate how Disinformers employed the chatbot to generate and validate information as part of their disinformation strategy.

\textbf{Generating Disinformation.}
\cattlehug{} and \floorlight{} used the chatbot to generate false information or fact-check real information to twist for disinformation purposes. 
\cattlehug{} sought to shorten statements or provide counterarguments, likely aiming to craft convincing disinformation while associating the Disinformer role with a high frequency of responses to push their agenda.
In contrast, \floorlight{} noted that they could have been underutilising the chatbot, attributing this to not being a \dquote{trained rumour monger}.

\textbf{Treating Chatbot as Strategist.}
\stampjab{} treated the chatbot as a strategist throughout the game, using prompts like, \dquote{I am playing a game similar to werewolf or mafia and have been given the role of a disinformer. I need to ensure that this role stays hidden from all other users...I need to blend in well with general public users so that they do not see me as trying to sway their opinions to mine. I will need your full support...}
\stampjab{} often used the chatbot to assist in writing comments, providing context on others' responses and the game happenings, as well as to generate true information that they could tweak slightly to become false.
During the post-game interview, \stampjab{} reflected that the chatbot helped ease some of their guilt by acting as a \dquote{neutral party}, making it easier for them to play their role. 
They also appreciated the chatbot's efficiency and speed in generating information, saving time compared to web searches, keeping them focused on the game. 
This was especially helpful as \stampjab{} did not personally agree with the stance they had to push but needed to persuade others to adopt it. 
Overall, \stampjab{} felt that the chatbot had been helpful and acknowledged it in helping them conceal as the Disinformer.
Similarly, \floorlight{} prompted the chatbot with \dquote{how to accuse someone of a role}.

\textbf{Getting Background Information and Mitigating Risk of Detection.}
\mailchart{} used the chatbot for purposes such as gathering background data on the topics, but also used the web to verify false information that they created as a precaution in case other participants attempted to expose them. 
Conversely, \ironbush{} used the chatbot to mitigate the risk of detection, typically using it to fact-check the disinformation that they wanted to share.
\ironbush{} also used the chatbot in coming up with examples that were used to provide substance to their responses to push the Disinformer agenda.
\ironbush{} even crafted prompts like \dquote{Craft a heartwarming message on why education is not the pathway to success}, which could have been aimed at using emotional content that plays on the empathy of others to push their agenda. 

These use cases highlight a dual purpose for the chatbot, where it could be used both in generating disinformation and also ensuring that the disinformation would be able to sustain scrutiny. 

\subsubsection{LLMs as Moderation Tools}

\begin{figure}[!htb]
    \centering
    \includegraphics[width=\linewidth]{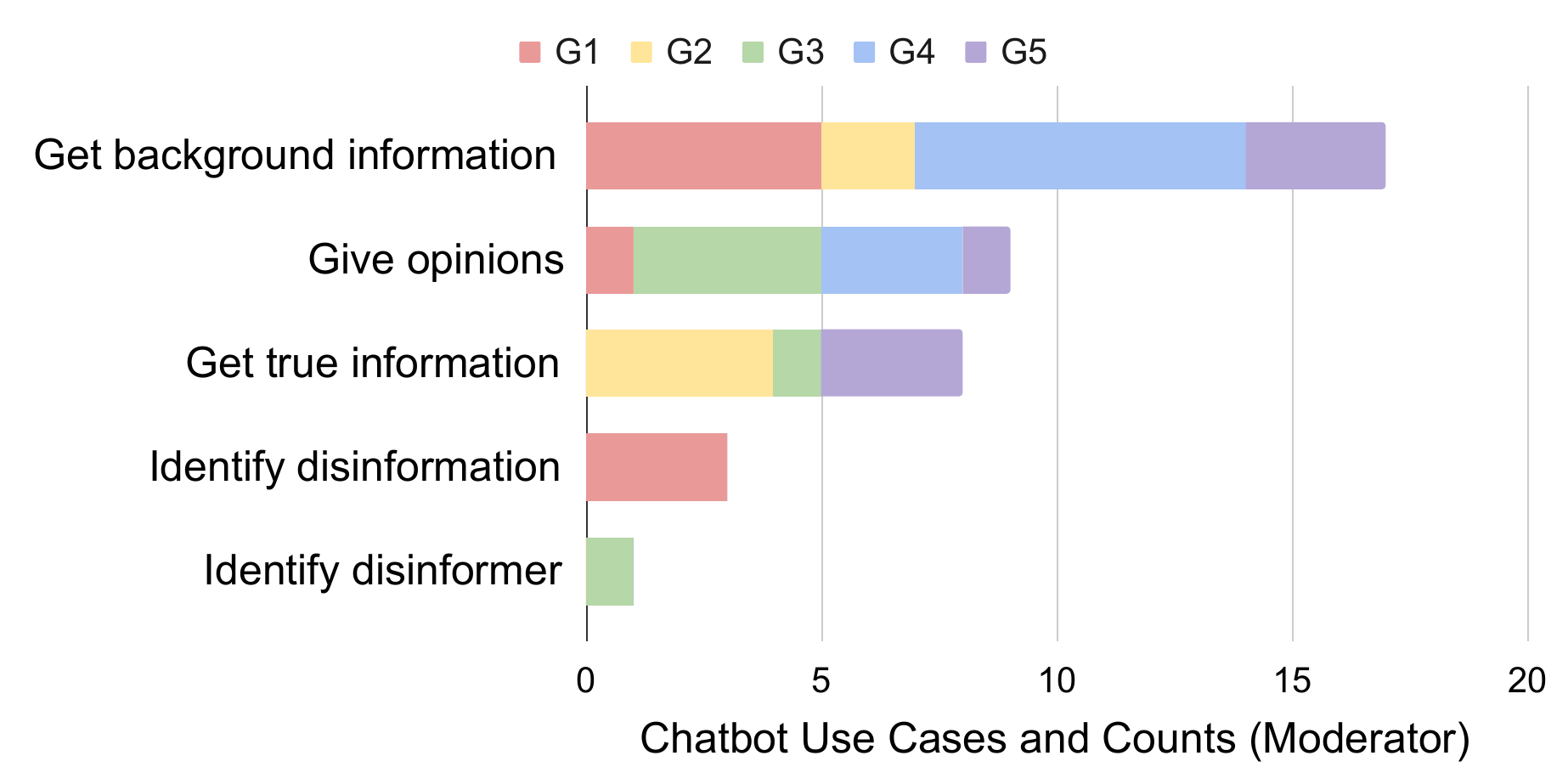}
    \caption{Chatbot use case counts for the 5 Moderators across all groups (\groupone-\groupfive). Counts range from 0 to 20.}
    \label{fig:mod_usage}
    \Description{A bar chart showing the distribution of chatbot usage by Moderators. The x-axis represents the frequency of chatbot usage, while the y-axis categorises five different purposes for using the chatbot. The legend at the top uses different colours to represent each of the five groups. Each colour-coded bar indicates how frequently the chatbot was used for a specific purpose in each group.}
\end{figure}

\begin{figure}[!htb]
    \centering
    \includegraphics[width=\linewidth]{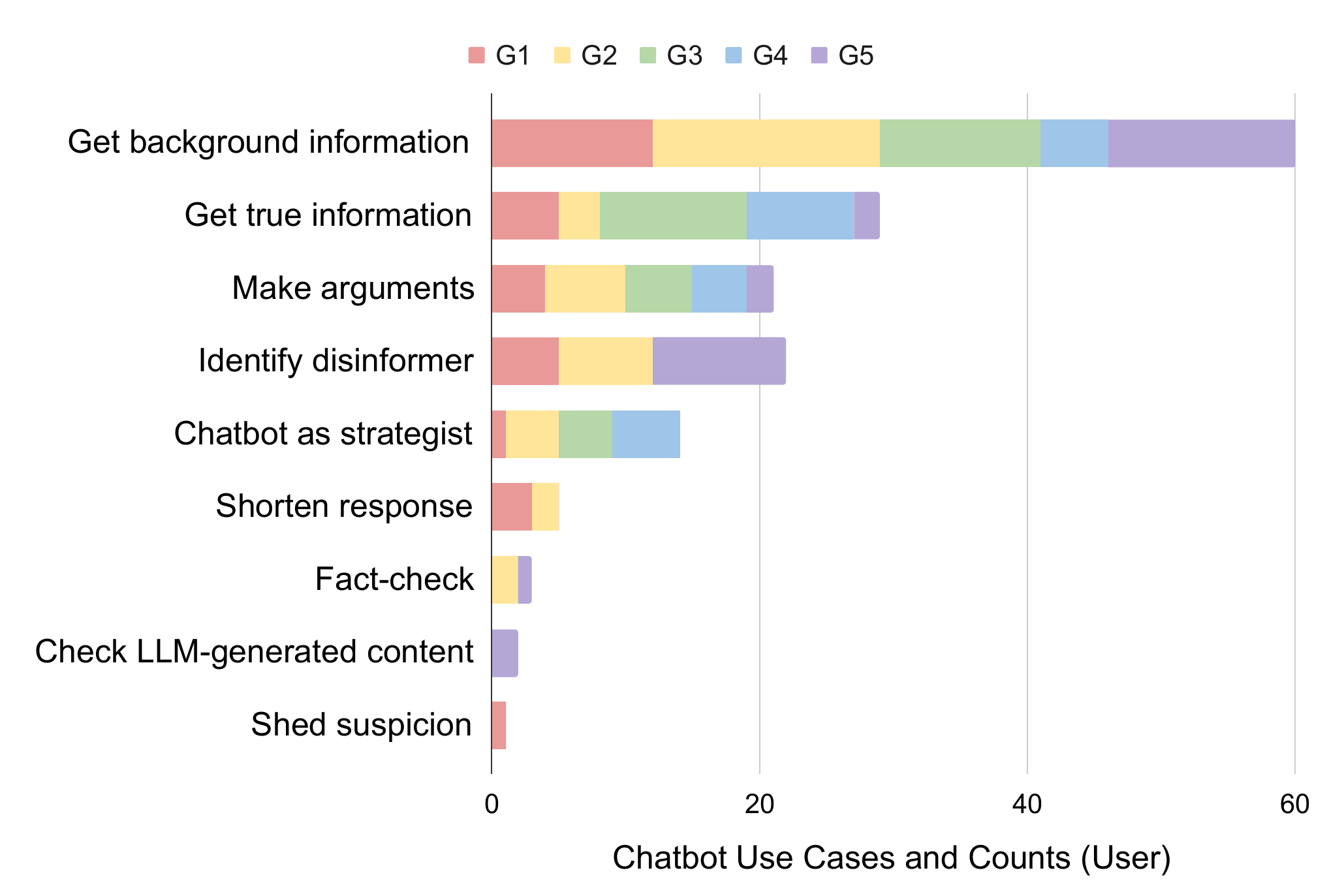}
    \caption{Chatbot use case counts for the 15 Users across all groups (\groupone-\groupfive). Each group has 3 Users. Counts range from 0 to 60.}
    \label{fig:user_usage}
    \Description{A bar chart showing the distribution of chatbot usage by Users. The x-axis represents the frequency of chatbot usage, while the y-axis categorises nine different purposes for using the chatbot. The legend at the top uses different colours to represent each of the five groups. Each colour-coded bar indicates how frequently the chatbot was used for a specific purpose in each group.}
\end{figure}

Chatbots served as fact-checking and content-verification tools for Moderators and Users, who monitored content and identified and challenged disinformation respectively.
The bar charts in Figures~\ref{fig:mod_usage} and~\ref{fig:user_usage} illustrate the distribution of chatbot usage by the Moderators and Users across all groups (\groupone-\groupfive). 
Moderators predominantly used the chatbot for getting background information ($n = 17$, $45\%$) and giving opinions ($n = 9$, $24\%$), often to verify content or contextualise discussions. Moderators also used the chatbot to identify potential disinformation ($n = 3$, $8\%$) and disinformers ($n = 1$, $3\%$).
Users also had a focus on gathering background information ($n = 63$, $43\%$) and true facts ($n = 27$, $18\%$), with less frequent use for identifying disinformers ($n = 16$, $11\%$), particularly in \groupone, \grouptwo, and \groupfive. These distinct uses influenced group dynamics, with the greater proportion of Users per session (1 Disinformer, 1 Moderator, and 3 Users) explaining the higher numbers of Users contributing to the broader chatbot interactions. 

\textbf{Getting Background Information.}
\plotsnail{} relied on the chatbot for neutral opinions and context on the topics, trusting its ability to offer unbiased insights. 
They often asked specific questions aligned with real-world issues and opinions held by others, while providing elaboration and context of the forum's discussion in their prompts where needed, as they believed the chatbot \dquote{might give a more unbiased opinion so that anybody, irrespective of their opinions, if they look at it, they might have some neutral opinion}. 

\stewmeter{} primarily used the chatbot to gather information and verify facts, often cross-referencing with the web, indicating a preference for traditional fact-checking for accuracy.
\shellhusk{} adopted similar reasons for using the chatbot cautiously, frequently cross-checking its output with external sources, indicating a general distrust of its reliability. They occasionally asked questions that were not directly related to the topic (e.g., \dquote{anything wrong with taxing the rich to help the poor?}), and occasionally requested for elaborations on possible stances for discussion (e.g., \dquote{why euthanasia should not be legalised}). 
Despite finding the chatbot repetitive and overly general, \shellhusk{} appreciated its ability to help \dquote{see things from a different perspective}, and noted that the repetitiveness might have stemmed from how they phrased their prompts.

\glidemail{} used the chatbot to fact-check and cross-reference statements, often engaging it for background information.
\lemonfret{} used the chatbot to ask questions, verify facts, and get background information, asking questions based on the discussion topic, such as \dquote{Do you think education is the pathway to success?}
However, \lemonfret{} found the chatbot less useful for deeper analysis, relying on it only for straightforward queries. This contrasted with other participants who used it more strategically.
\pigwhisper{} primarily used the chatbot to ask for summaries or fact-checking and was more focused on verifying information through multiple sources rather than heavily relying on the chatbot. 

\textbf{Identifying Disinformer and Disinformation.}
\plotsnail{} and \lemonfret{} asked the chatbot to analyse comments from suspected Disinformers to determine if they aimed to \dquote{sway the public opinion towards his agenda}. 

In contrast, \subtlerun{} directly asked \dquote{Who do you think is the disinformer this round?}, without providing gameplay context. This suggests they may have expected the chatbot to be context-aware and capable of understanding the ongoing discussion in the forum. Their less consistent usage of the chatbot, with generic prompts like \dquote{What is success?}, suggests an assumption that the chatbot could infer game details without the necessary information.

\pigwhisper{} provided the entire content of the discussion forum but disregarded correct suggestions in favour of their own choices.
\sunnygoat{} used the chatbot to identify LLM-generated content, associating such responses with disinformation.
\tribeboat{} and \gelpick{} also provided it with gameplay details to identify the disinformer but only achieved partial success.

Instead of sticking to basic fact-seeking queries, \jewelpipe{} asked the chatbot strategic questions, such as \dquote{What sort of commands would you pick in order to suss out a distractor to a constructive conversation?} They acknowledged critical evaluation and acknowledged the chatbot’s limitations, noting that \dquote{evidence is needed for chatbot responses to be valid}. 

\textbf{Limited Use Due to Prior Knowledge.}
\subtlerun{} used the chatbot less consistently, primarily consulting it early in the game. In their post-game interview, they explained that they felt \dquote{quite familiar with the topic} and \dquote{didn't have to rely on the chatbot so much.}
\webcube{} regarded the chatbot as a supplementary tool rather than a necessity, relying primarily on prior knowledge. They expressed reservations about depending solely on the chatbot, using it mainly to confirm facts or gather information that aligned with their existing knowledge or background. For example, they asked questions like, \dquote{What are some of the social welfare schemes provided for \redact{a country's citizens}{Singaporeans} as of 2022?}
They mainly sought to reaffirm facts they already believed to be true, noting that their prompts were \dquote{mainly just to check if the AI could provide [them] the answer based on the assumption of what [they] thought it was capable of.}
Notably, they suggested that the arguments and topics \dquote{were mostly general} and had \dquote{extremely sweeping statements that [they] had to verify.} They added that if they were already familiar with the topic, they \dquote{would not want to use the chatbot at all}, doubting it would offer new insights.

These use cases highlight that the chatbot served as both a helpful tool and a source of scepticism. Its role in aiding moderation and content verification varied, shaping the dynamics of each group.

\subsubsection{Limitations of LLM Observed by Participants}
While participants found the chatbot useful, they also highlighted several limitations.

\textbf{Verbosity and Hallucination.}
\stampjab{} frequently asked the chatbot to shorten responses and make them sound more natural, with prompts such as \dquote{Keep in mind my goal, could you draft a common language reply to that question?} or requests to add concrete examples for clarity.
Ironically, when \stampjab{} tried to generate statistics that could be manipulated into disinformation, the chatbot hallucinated and provided an incorrect statistic, stating that life expectancy in 2021 was \dquote{8.0 years for males and 8.5 years for females}. When they attempted to tweak the year to make the statistic appear false, they inadvertently corrected the error, posting accurate data instead: \dquote{80 years for males and 85 years for females in 2023}.
Despite these limitations, they believed that the chatbot helped in concealing their identity.
 
\textbf{Conspicuous Disinformation Generation.}
\mailchart{} failed to use the chatbot to create convincing falsehoods that would appear reasonable and be difficult for others to dispute, finding that while it could help identify information gaps, it often produced exaggerated news articles that would have been easily recognised as false when checked. This likely occurred because \mailchart{} had asked the chatbot to produce \dquote{fake news}, leading to overly obvious results. As a result, they manually created their own disinformation instead.

\textbf{Lack of Human Touch.}
\bootspark{} pointed out that the chatbot sometimes produced responses with a generic or artificial tone if not carefully edited.
\sunnygoat{} noted the chatbot's \dquote{lack of human touch}, preferring shorter, more natural-sounding responses. They felt that overly formal replies could appear LLM-generated, raising suspicion of disinformation.

\textbf{Unhelpful Chatbot, Alternatives Preferred.}
\midfrost{} found the chatbot unhelpful for role-related tasks, expressing a dislike towards it for \dquote{using it to come up with ideas or anything like that}, instead only using it to \dquote{meet requirements} and \dquote{clarify things}. They also emphasised that they \dquote{would have rather gone to Google}, as they believed the web was better for specific, reliable information or facts.
\hollowsea{} felt that the chatbot was useful for generating general talking points and surface-level information, but recognised its limitations after receiving incorrect information and concluded that while it could provide a helpful overview, it should not be relied on for important decisions.

\subsection{Comparative Analysis of LLM and Web Search Usage} 

\textbf{Fact-Checking.}
Some participants preferred the web over the chatbot for verifying facts or obtaining detailed information, especially when they had specific outcomes in mind and trusted the web for complete and credible sources like news articles or official statistics. \lemonfret{} and \bootspark{} relied on the web for fact-checking, as they found it easier to access complete, credible information—something they felt the chatbot could not always provide. Specifically, \bootspark{} preferred web browsers in checking official sources, as it allows access to complete information, such as news articles or official statistics, which could not be fully verified through the chatbot alone. \stampjab{} also believed web searches offered a broader perspective.

Concerns about the chatbot's reliability were common, with several participants noting its tendency to hallucinate or provide inaccurate, incomplete information. \jewelpipe{} remarked that it would be \dquote{embarrassing to spout nonsense} while \gelpick{} highlighted issues with contextually irrelevant information, such as the chatbot providing content from a different country to what was asked.
Similarly, \midfrost{} noted that the chatbot might not provide the details available through web searches. 
\hollowsea{} expressed concern that while chatbots aggregate information, they do not provide reliable sources, potentially diminishing trust due to the hallucinations in such AI-generated content, as opposed to a web browser that would present the sources to the user easily.

Notably, \tribeboat{} and \hoodclad{} adopted a hybrid approach in using both tools, with the chatbot used for getting general overviews and web searches used for fact-checking and gathering reliable data to substantiate their posts.

\textbf{Get Background Information.}
While participants generally preferred the web over the chatbot for fact-checking, their preference for gathering background information was less consistent.
Some preferred the web, with \glidemail{} valuing its ability to provide comprehensive overviews and \hallpaste{} using it to source for information to subsequently feed into the chatbot.
\cattlehug{} noted that the web offered multimodal data (e.g., images and videos), whereas the chatbot only provided text.
Conversely, \plotsnail{} found the web overwhelming due to the abundance of available information and preferred the simpler, faster responses of the chatbot.
\sunnygoat{} noted that the chatbot was useful for listing actionable ideas, searching for information, and brainstorming content. 
\midfrost{} highlighted the chatbot's usefulness for more conceptual topics (e.g., \dquote{the role of education in success}), where the content might not be readily available through web searches.
\section{Discussion}

We examine our findings through the lens of the current capabilities and challenges associated with using LLMs in both escalating and mitigating disinformation. We also discuss the varying levels of reliance and scepticism exhibited by participants toward LLM-generated outputs. Finally, we seek to draw insights that can inform future research, policy decisions, and the design of safeguards within online social platforms. 

\subsection{The Sword and the Shield: Dual Roles of LLMs in Elevating and Combating Disinformation}

\subsubsection{Sword: LLMs to Exacerbate Disinformation}

LLMs can support Disinformers by helping them generate content to boost their agenda and providing strategic guidance towards being a more efficacious Disinformer.
In our study, Disinformers tended to wield the chatbot as a \textit{sword}—creating and amplifying disinformation.
Disinformers used the chatbot more frequently, partly due to the need to manipulate and produce content aligned with their agenda.

\textbf{Produce Disinformation.}
LLMs can generate large amounts of misleading or false information in various forms.
For example, \cattlehug{} and \floorlight{} used the chatbot to produce falsehoods, verify their own fabricated content, and gather background data that could be twisted into misleading narratives. 
The chatbot provided quick, context-aware responses, which could be blended with factual data or slightly tweaked to create disinformation. It also helped Disinformers create strategic arguments and counterarguments, bolstering their ability to push their agenda and control narratives.
Such uses of LLMs have been corroborated by existing studies that found that while LLMs can offer immense support in generating content, they also pose risks, particularly when used to craft misleading narratives~\cite{borji2023categorical, zhuo2023red}. 
Malicious actors
can weaponise such models to automate disinformation, allowing them to be spread on a larger scale~\cite{barman2024dark}.

\textbf{Disinformation Strategy Advisor.}
We observed that Disinformers not only leveraged the chatbot to produce disinformation but also to strategise how to advance their agenda effectively, positioning the chatbot as a strategic advisor.
The chatbot provided suggestions on how to avoid detection, such as \dquote{disinformation tactics, language analysis, social engineering, misdirection, information management and presentation}.
When asked how one can avoid being suspected, the chatbot suggested that the Disinformer could try strategies such as staying vigilant and observing other participants' behaviour and communication patterns. 
By employing the LLMs in this manner, the Disinformers could refine their strategies and make it more difficult for others to expose them. 
For instance, \floorlight{} prompted the chatbot with \dquote{how to accuse someone of a role}, hoping to push the suspicion onto another participant. 
\stampjab{} also often asked the chatbot for ways to remain hidden as the Disinformer.

\subsubsection{Shield: LLMs to Mitigate Disinformation}
LLMs can support moderators and users by enhancing their contextual knowledge and verifying content. They can also provide strategic guidance towards identifying the Disinformer and shaking off suspicion.
In our study, Moderators and Users, on the other hand, approached the chatbot as a \textit{shield} to defend against disinformation. The chatbot helped to fact-check content and identify potential disinformation, rather than generating content.

\textbf{Content Verifier.}
Participants used the chatbot to fact-check and cross-reference claims made by other participants.
For instance, \glidemail{} used the chatbot to expose potential disinformation by cross-referencing participants' statements with verified facts. 
This approach was used to identify discrepancies or falsehoods in the arguments presented by others, particularly those they suspected of being the Disinformer.
Existing studies have found that LLMs have diverse use cases
such as searching for information, data analysis, answering questions, and providing informational support and advice~\cite{dam2024complete}.
Chatbots have also been used to support fact-checking efforts by allowing users to verify information from reliable sources in real time by disclosing their chat content directly to fact-checking services~\cite{lim2023fact}, and seeking trustworthy information on specific topics~\cite{gupta2021truthbot}.

\textbf{Guide Mitigation Approach.}
Beyond fact-checking and providing information, the chatbot provided strategic guidance to Moderators and Users in identifying the Disinformer, providing ways to draw out the Disinformers, or even to avoid being suspected as the Disinformer.
\pigwhisper{}, \tribeboat{} and \gelpick{} used the chatbot to identify the Disinformer with varying degrees of success.
The presence of the chatbot also provided participants with novel strategies and avenues to approach forum interactions, such as refining their posts, bolstering their arguments, seeking advice on ways to engage, and even steering group dynamics.
\tunakept{}, who faced suspicion due to a minority stance, actively used the chatbot to generate arguments and refine their approach to avoid further suspicion. 
They asked the chatbot for advice on persuading others of their innocence, seeking strategies that might help them blend in better with the group. 
This proactive chatbot use demonstrates how participants recognised its potential for strategic guidance.

\subsection{Obstacles to Effective LLM Use}
Our study surfaced several obstacles impeding participants' effective use of LLMs, arising from both LLMs' inherent limitations and user biases and behaviours. Additionally, scepticism toward LLM-generated content shaped how outputs were evaluated and utilised. Research suggests that an exclusive focus on building trust in AI may be unwarranted, as chatbots can produce convincing but false information~\cite{peters2023importance}. 

\subsubsection{\textbf{LLM-Based}: Verbosity and Formality}
Participants found the chatbot’s responses overly verbose and formal, making them unsuitable for casual online forums and easily identifiable as AI-generated. This made participants hesitant to rely fully on the chatbot, echoing findings from Fu et al.~\cite{fu2024text}, who noted that chatbot outputs often tended to be unnaturally long and filled with irrelevant tangents, which is problematic in informal, high-stakes communication environments. In our study, participants either selected useful portions of the chatbot output or discarded them entirely.

This verbosity also posed a unique risk for Disinformers, as seen when \cattlehug{} asked the chatbot to include false information and the response included an explicit disclaimer about the fabrication. Had they not carefully edited the output, they could have been exposed.

\subsubsection{\textbf{LLM-Based}: Knowledge Cutoff and Hallucination}
Another limitation was the LLM’s knowledge cutoff, which restricted its ability to provide up-to-date information. While the cutoff date for both our model and its base model is not officially reported, Chatbot Arena, a platform for evaluating LLMs, suggests it to be January 2024~\cite{chiang2024chatbot}. However, even official cutoff dates often differ drastically from effective ones~\cite{cheng2024dated}. The fine-tuning dataset used to create our study’s LLM largely comprises synthetic data generated by GPT-4, which has an earliest reported cutoff of September 2021~\cite{OpenHermes-2.5}. Therefore, the effective cutoff could range from September 2021 to January 2024, meaning the model may lack reliable information on events after September 2021.

Coupled with occasional hallucinations—where the chatbot generated inaccurate information—this reduced the chatbot’s reliability as a fact-checking tool. For instance, participants noted instances where the chatbot provided outdated or incorrect information, further eroding trust in its outputs. When \hollowsea{} sought the Gross Domestic Product of \redact{a country}{Singapore} from the chatbot, they obtained a response of \dquote{Gross Domestic Product (GDP) for the year 20 was US\$36.3 billion}, which they identified to be incorrect: \dquote{Okay, chatbot being wrong. It's wrong by a factor of 10.} Following up with with \dquote{what is this year}, the chatbot responded \dquote{As of my last update on this topic, the current year is 2022.}, undermining its credibility and potentially causing scepticism. 

\subsubsection{\textbf{LLM-Based}: Lower Capabilities of Open-Sourced LLMs}
The reasoning abilities of the LLM used could have also limited participants’ ability to obtain useful responses. While uncensored open-source models like the one used in our study~\cite{OpenHermes-2.5} are valuable for research, they lack the robust reasoning abilities and up-to-date information available in commercial models like GPT-4~\cite{achiam2023gpt}, Claude 3 Opus~\cite{anthropic_claude-3}, and Gemini 1.5~\cite{reid2024gemini}, which also incorporate safety guardrails to mitigate misuse.
Though this model is state-of-the-art for its size and matches or outperforms previous closed-source models like GPT-3.5 on many benchmarks~\cite{jiang2024mixtral}, its reasoning capabilities still fall short of the best closed-source models available at the time. As a result, the overall experience and usability of the chatbot for our study were impacted.

Nevertheless, the presence of guardrails on the best LLMs gives well-meaning users an advantage over ill-intentioned users, as they have access to the best-performing tools, particularly since it is in the interests of the companies to ensure that their models are always top-performing. Malicious actors, however, are often limited to uncensored yet subpar models, which are typically outdated and less effective in producing contextually relevant disinformation on current affairs. They may consider alternatives like jailbreaking and bypassing the guardrails of commercial LLMs, but even a successful attempt may quickly be disrupted given the vigilance of the companies on such misuses~\cite{InstructionHierarchyTraining}. Alternately, they could build their own LLM, but this approach is neither cheap nor straightforward~\cite{WhatLargeModels}—likely feasible only for the most well-funded of bad actors.

\subsubsection{\textbf{User-Based}: Selective Trust and Critical Engagement with AI-Generated Content}
Participants’ trust in the chatbot’s outputs was shaped by their existing knowledge and perception of its accuracy. Those confident about the topic often used the chatbot to reinforce existing views rather than challenge them, reducing the chatbot's potential in facilitating discovery and critical engagement. Opinionated LLMs could exacerbate this by reinforcing such biases~\cite{10.1145/3613904.3642459}.
Those who doubted the chatbot’s accuracy were more likely to critically evaluate its outputs instead.

Content that were thought to be AI-generated also garnered scepticism, though to a lesser extent.
\cattlehug{}'s posts were suspected of being AI-generated, but were still tolerated for several rounds, arousing strong suspicions only in the last round, suggesting that the suspected involvement of AI alone was not enough to trigger distrust or rejection of the content. Other factors, such as the relevance of the post (e.g., \floorlight{}'s off-topic statements) or social behaviour of participants (e.g., \gelpick{}'s emotional outbursts), took precedence.
On the contrary, when participants openly declared that their posts were written with the use of AI (e.g., \jewelpipe{} copied and posted \dquote{from AI; Sure, I will help with that. "Military life" refers...}), such posts were accepted, receiving upvotes.
This suggests a baseline level of trust in AI-generated content when transparency is present.

These observations raise key questions about LLMs in environments with users of varying expertise. Knowledgeable participants may rely on chatbots as a confirmatory tool, overlooking personal errors or biases, while less knowledgeable participants may trust AI content, making them vulnerable to potential hallucinations in them. Future research should explore these pitfalls, examining how users with different knowledge levels engage with and critique AI outputs to design AI systems that foster critical engagement and mitigate reinforcing cognitive biases.

\subsubsection{Implications for Policymakers and Developers in Navigating LLM-Assisted Disinformation}
The identified limitations provide guidance for policymakers, AI developers, and industry stakeholders on managing LLMs in the context of disinformation on social platforms, highlighting their dual role as both tools and barriers.
\begin{itemize}
\item \textbf{Differential Access and Safeguards}: Limit access to top LLMs and strengthen safeguards to prevent misuse. The gap between commercial and open-source models helps curb disinformation, while robust guardrails and legal frameworks~\cite{UsagePolicies, GenerativeAIProhibited} can offer added protection.
\item \textbf{Enhancing User Literacy}: As LLMs become integrated into social platforms~\cite{MeetYourNew2024}, public education initiatives are essential to empower users to critically assess AI-generated content. Findings from our study show that scepticism toward AI content is influenced more by content relevance and social dynamics than mere suspicion of AI involvement. This aligns with recent research~\cite{buchanan2024people} that highlighted the context-dependent nature of user trust in AI-generated content. Equipping users with tools and training to evaluate outputs can reduce biases, prevent reliance on simplistic prompts, and help users harness LLMs effectively for information verification~\cite{chen2023unleashing}.
\item \textbf{Cross-Sector Collaboration}: Promote partnerships across stakeholders including policy-makers, media firms and tech companies to address LLM-enabled disinformation, balancing innovation with safeguards and addressing the interplay of legal frameworks, technology, and user behaviour.
\end{itemize}

By prioritising these areas, policymakers and developers can leverage LLMs to counter disinformation, mitigate risks of misuse, and adapt to the evolving landscape of online information dissemination. Labels for AI-generated content, tools for critical engagement, regulations ensuring AI accountability, and fostering digital literacy are essential to navigate the evolving challenges of integrating AI into digital systems are susceptible to disinformation.

\subsection{Designing for Appropriate Reliance on LLMs}
Insights from our study highlight the need to balance trust and scepticism to enable constructive and informed engagement with AI-generated content. Appropriate reliance on AI systems is crucial for their effective use and oversight, especially given the increasing call for human oversight to safeguard against AI failures. A key challenge lies in preventing overreliance on AI, which is defined as \dquote{users accepting incorrect AI outputs, which can lead to errors and ultimately erode trust in these systems}~\cite{passi2022overreliance}.

Many participants showed scepticism toward the chatbot’s outputs, such as by choosing not to use the chatbot’s content directly (\sunnygoat{}), while some frequently refined LLM-generated responses, particularly when crafting disinformation (\mailchart{}, \ironbush{}). Many also cross-checked chatbot outputs against other sources, such as web browsers, to ensure accuracy. Explicit distrust was voiced by participants who noted receiving incorrect information from the chatbot (\hollowsea{}). Moreover, most participants relied on the chatbot for low-stakes tasks, such as brainstorming or gathering background information. Successful Disinformers exercised greater control, carefully editing chatbot outputs, and performed better, receiving fewer suspicion votes (\mailchart{}, \ironbush{}, \stampjab{}). In contrast, less successful Disinformers relied more heavily on unrefined chatbot content and were easier to detect or were off-topic (\cattlehug{}, \floorlight{}).
There were also instances of strong reliance on the chatbot, such as users using it to identify the Disinformer (\gelpick{}) and directly copying LLM-generated content into discussions (\hallpaste{}, \hoodclad{}). While these cases suggest implicit trust in the chatbot, they were relatively rare and not indicative of broader behavioural patterns.

These findings underscore the importance of fostering appropriate reliance on LLMs. Future research and policy considerations should aim to enhance users' critical evaluation skills when interacting with LLMs, particularly in contexts with a high potential for disinformation. Additionally, LLM developers can consider implementing features, e.g., slow algorithms that impose a waiting time to raise analytical thinking~\cite{10.1145/3359204, 10.1145/3512930}, that encourage users to critically assess and verify LLMs-generated outputs, especially for tasks with significant real-world implications.

\subsection{Implications for Future Design of Online Social Platforms}

Our study reveals chatbots' dual role in propagating and combating disinformation in online forums. Disinformers used it to generate misleading content, plan disinformation strategies, and influence discussions, while others leveraged it to verify facts, strategise disinformation detection, and engage constructively. However, participants also expressed concerns about verbosity, outdated information, and hallucination, contributing to a general lack of trust in it.

As a precursor to outlining future developments, it is important to consider key ongoing advancements in LLM research:
\begin{itemize}
    \item \textbf{Human-Aligned Fine-Tuning}: Aligning LLMs with traits like helpfulness, honesty, and harmlessness~\cite{bai2022constitutional}.
    \item \textbf{Retrieval-Augmented Generation (RAG)}: Enhancing accuracy and reducing hallucination of LLMs by incorporating external databases as context~\cite{gao2023retrieval}.
    \item \textbf{Autonomous AI Agents}: Systems like AutoGPT~\cite{yang2023auto}, with access to external tools, and capable of autonomous planning, reasoning, and execution of goals.
\end{itemize}

As these technologies continue to evolve, we foresee potential threats regarding the misuse of LLMs in online platforms:
\begin{itemize}
    \item \textbf{Malicious Fine-Tuning}: Techniques like abliteration~\cite{UncensorAnyLLM} and fine-tuning on malicious data~\cite{GPT4Chan2024} could enable models to generate disinformation and hate speech easily.
    \item \textbf{Malicious RAG}: Bad actors could create databases of harmful context or skewed narratives to improve the planning and execution of disinformation campaigns.
    \item \textbf{Malicious AI Agents}: While currently limited in effectiveness~\cite{MeetChaosGPTAI}, automated AI agents designed with destructive intent may evolve to plan and orchestrate large-scale disinformation campaigns.
\end{itemize}
To mitigate these threats, future platforms should employ AI tools that encourage prosocial behaviours while restoring trust in AI systems. While progress is being made with AI interventions that foster consensus and encourage critical evaluation of machine-generated content~\cite{govers_ai-driven_2024, gould_chattldr_2024}, more can be done. Potential solutions include:
\begin{itemize}
    \item \textbf{Personalised Intelligent Chatbots for Users and Moderators}: Platforms can integrate fine-tuned chatbots that can understand forum context and reliable online knowledge databases for RAG. Fine-tuned to reason about stances, current affairs, and social engagement strategies, they should ensure harmlessness and resistance to jailbreaks. With explicit user consent, they could also adapt to users’ communication style, balancing content engagement with disinformation detection, while allowing users to control the level of AI involvement.
    \item \textbf{AI Content Detectors and Disinformation Flaggers}: Automated AI content detectors to flag potential AI-generated text and disinformation, helping users distinguish between human and AI-generated content, fostering transparency and informed decisions about AI involvement.
    \item \textbf{Post-Augmentation with AI Insights}: LLMs can enrich content with information on the likelihood of it being AI-generated, stance alignment, automated fact-checking and source attribution. Balancing automation with human oversight increases transparency, trust, and critical engagement with the content.
\end{itemize}

However, such measures may create tension between the human desire for self-expression and the need for factual accuracy~\cite{MeetChaosGPTAI}. As such, it is crucial to address concerns about implicit censorship and the potential risk of stifling opinions when encouraging a landscape that prioritises factual and verified information.
\begin{itemize}
    \item \textbf{Preserving Space for Opinion and Subjectivity}: Overreliance on AI content detectors and disinformation flaggers risks censoring opinions, especially in subjective discussions. AI tools must distinguish between factual verification and subjective expression, only flagging factual inaccuracies or disinformation without suppressing opinions or subjective interpretations unless they are explicitly harmful or misleading.
    \item \textbf{Adaptive Fact-Checking and Content Moderation}: AI moderation should be context-sensitive. Light moderation would suit opinion-driven discussions as factual accuracy is less emphasised, while scientific or political topics might require more stringent fact-checking, by using approaches such as the one suggested by Sehat et al.~\cite{sehat_misinformation_2024}. This flexibility allows diverse discussions without stifling opinions.
    \item \textbf{Transparency and User Empowerment}: AI tools must be transparent to avoid censorship perceptions. Clear explanations for flagged content and the option to challenge or appeal unfairly targeted opinions can help to maintain trust and user control, especially regarding concerns about being confined to a fact-first environment.
\end{itemize}

These designs and safeguards for future platforms aim to empower both users and moderators with AI tools that promote critical thinking, transparency, and trust. By integrating AI systems with careful safeguards, online forums can minimise disinformation while preserving authentic, prosocial human interaction at the core of online engagements.
\section{Limitations}

With this study being conducted as a role-playing game, a inherent limitation is that the Disinformers and Moderators were non-professional actors--limited by their abilities to play out their roles--since real-world disinformers are not practically available.
Nonetheless, our study maps out the space in which LLMs can potentially be used to advance and combat disinformation, laying the groundwork for future research.

Participants had a checklist of actions during the game and were given reminders to complete them to ensure a baseline level of interactions.
They also occasionally consulted the researchers on how to complete the tasks.
While such interactions may reduce the authenticity of participants' behaviours, we referred them to the influence guide they had to minimise such occurrences.

While we measured participants' stances towards the topics before and during the game, we saw only minute shifts towards the direction of the Disinformer's agenda (0.25 to 1 on a 7-pt Likert scale in 3 out of 8 instances), suggesting that opinions remained largely unchanged. Given that opinions are known to be resistant to change~\cite{pomerantz1995attitude}, future research could explore the long-term effects of disinformation on opinion flexibility and investigate how repeated exposures to disinformation might gradually shift attitudes over time.

Our work shows that a small-scale game setting can elicit disinforming and defensive behaviours with LLMs. Future work could expand the game design towards a larger scale to explore more complex behaviours such as multiple disinformers colluding and how others counteract these coordinated actions.

The creation of disinformation using large multimodal models (LMMs) should also be considered. While this study focused only on LLMs--especially since text is a basic and cheap medium to produce disinformation, other mediums like images and videos can be more persuasive and believable~\cite{hameleers2020picture, gamage_deepfakes_2022, jaidka_misinfo_disinfo_genai_2024}. The space of LMMs in the context of disinformation remains largely unexplored and warrants closer examination.

\section{Conclusion}
This study provides a first-of-its-kind exploration of the role of LLMs in the disinformation scene, investigating their use in both spreading and countering such content in small, localised settings.
While prior work has utilised various methods to study disinformers' actions, it remains a challenge to get to know their true disinforming intentions.
Through our custom \textit{Werewolf} game simulation, we observed how Disinformers, Moderators, and Users leveraged LLMs in an online forum that uniquely provided insights into the intentions and strategies of motivated role-playing disinformers.
The findings reveal the dual nature of LLMs: while Disinformers effectively generated persuasive, misleading content, Moderators and Users often struggled to fully utilise LLMs for fact-checking and verification.
The broader implications suggest that, although LLMs have potential as tools to combat disinformation, their limitations—especially in detecting subtle manipulation—warrant further research. Improved safeguards and better AI training are essential to prevent their misuse. Additionally, educating users to critically assess AI-generated content is crucial, particularly on social media, where the distinction between human- and machine-generated content is increasingly blurred.
Future work should refine LLM tools for more effective real-time disinformation detection, explore their integration into social platforms, and assess the long-term societal impact of AI-driven content moderation. Addressing these challenges will help mitigate the risks of disinformation in an AI-powered world.

\begin{acks}
Our gratitude to Kohleen Tijing Fortuno for her assistance with the pilot studies.
\end{acks}

\bibliographystyle{ACM-Reference-Format}
\bibliography{main}

\onecolumn
\newpage

\appendix
\section{Appendix}
\begin{figure*}[!htb]
  \centering
  \includegraphics[width=.9\textwidth]{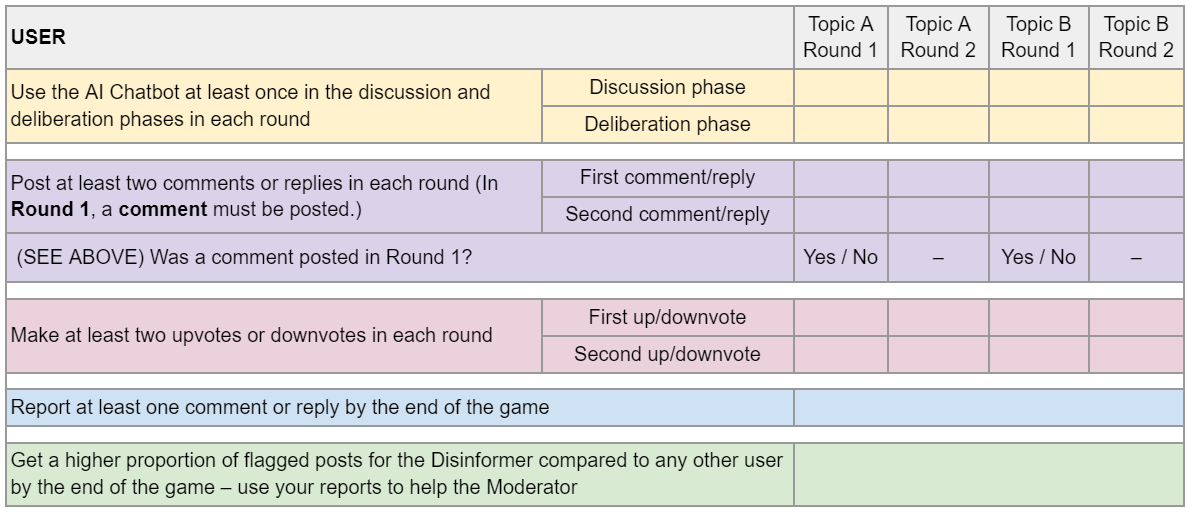}
  \caption{User's checklist. Each role has a checklist to ensure some common minimal interactions on the platform and role-related interactions to be completed during the game. For instance, the User is expected to post at least two comments or replies during each round.}
  \label{fig:checklist-user}
  \Description{The figure shows a checklist of tasks for the User engaging in forum discussions. For example, the User is required to use the AI chatbot, post comments or replies, make at least two upvotes or downvotes during each round, and report at least one comment by the end of the game.}
\end{figure*}

\begin{figure*}[!htb]
  \centering
  \includegraphics[width=.9\textwidth]{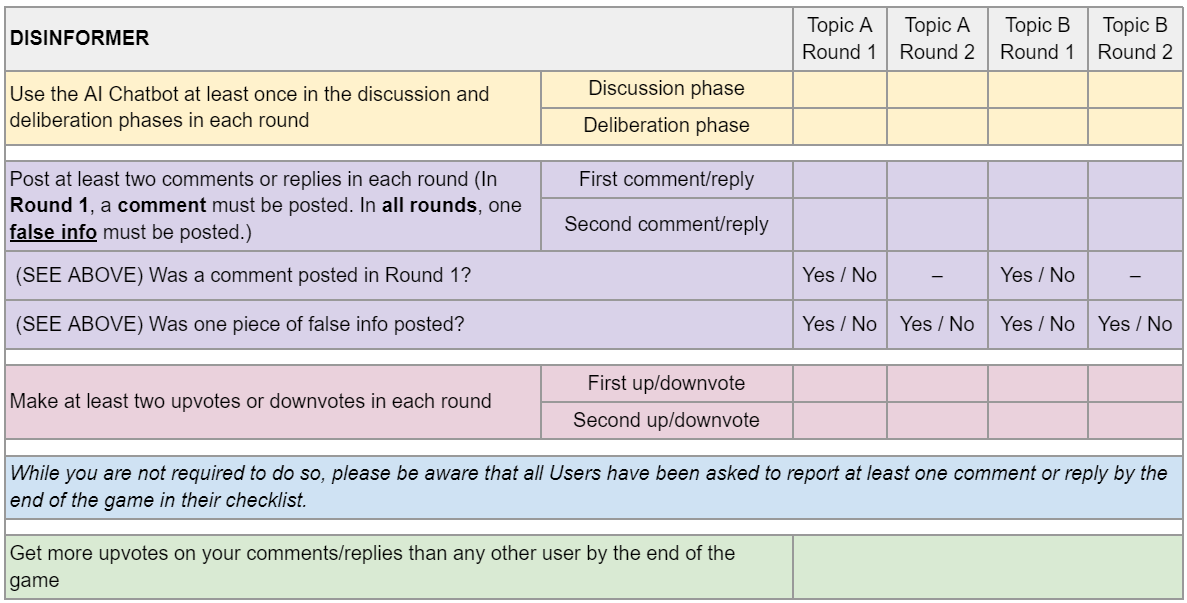}
  \caption{Disinformer's checklist. Each role has a checklist to ensure some common minimal interactions on the platform and role-related interactions to be completed during the game. For instance, the Disinformer is expected to post at least one piece of false information during each round.}
  \label{fig:checklist-disinformer}
  \Description{The figure shows a checklist of tasks for the Disinformer engaging in forum discussions. For example, the Disinformer is required to use the AI chatbot, post one piece of false information, make at least two upvotes or downvotes during each round, and get more upvotes on their comments or replies than any other user by the end of the game.}
\end{figure*}

Figures \ref{fig:checklist-user} and \ref{fig:checklist-disinformer} show the checklists of tasks that the Users and Disinformers have to complete, outlining their respective objectives.

\newpage

Table~\ref{tab:all_groups} shows the key events and group characteristics for the five groups.

{
\small
\begin{longtable}[!htb]{c p{0.5\textwidth} p{0.4\textwidth}}
\caption{Summary of key events and group characteristics for the five groups. Roles are denoted as follows: \textbf{D - Disinformer}, \textbf{M - Moderator}, \textbf{U - User}. For each of the four rounds in a game, disinformer suspects are listed based on the player(s) with the majority vote. In cases where the vote was tied, suspects are separated by a slash (/).}
\label{tab:all_groups}
\Description{A table that shows the key events and group characteristics of each group where the key events are mapped across the four rounds in a game and the group characteristics indicate the flow of the game.}\\

\toprule
Group & Key Events (R = Round) & Group Characteristics \\
\midrule
\endfirsthead

\multicolumn{3}{c}
{\tablename\ \thetable\ -- \textit{Continued from previous page}} \\
\toprule
Group & Key Events (R = Round) & Group Characteristics \\
\midrule
\endhead

\multicolumn{3}{r}{\textit{Continued on next page}} \\
\endfoot

\bottomrule
\endlastfoot
\groupone{}
& 
\begin{minipage}[t]{\linewidth}
  \textbf{R1:} A strong minority stance expressed by \tunakept{} sparks suspicions lasting almost the entire game. \cattlehug{} posts AI-generated content that goes largely unnoticed.

  \textbf{R2:} \tunakept{} defends their stance amid increasing suspicions. \cattlehug{}'s AI-generated posts are acknowledged but not fully recognised as disinformation.

  \textbf{R3:} \tunakept{} adapts by being less vocal. \cattlehug{}'s AI-generated disinformation starts raising suspicions.

  \textbf{R4:} \cattlehug{}'s AI-generated posts are finally recognised as suspicious, shifting focus towards identifying them as the disinformer.

  \textbf{Disinformer votes:} \tunakept{}; \tunakept{}; \tunakept{}; \cattlehug{}
\end{minipage} 
&
\begin{minipage}[t]{\linewidth}
  \begin{itemize}[leftmargin=*]
    \item Conventional group; less active participation but general adherence to roles and earnest in terms of interactions.
    \item Pivotal 1st round: \tunakept{}'s strong minority stance provided cover for \floorlight{}, who remained largely undetected early on.
    \item Despite initial misconceptions, the moderator and users showed adaptability, eventually shifting focus and correctly identifying the disinformer in the final round.
    \item Overall, the group adjustment and attentiveness overcame early misunderstandings to reveal the disinformer.
  \end{itemize}
\end{minipage} \\
\midrule
\grouptwo{}
& 
\begin{minipage}[t]{\linewidth}
  \textbf{R1:} \floorlight{} begins with an AI-generated post about controlling foreign worker numbers, drawing criticism and independent votes.

  \textbf{R2:} \floorlight{} makes a controversial statement about \redact{a locality}{Toa Payoh} being a \redact{non-local}{non-Singapore} town, attracting criticism and a flag from the moderator. They attempt to defend their remark amid growing scrutiny.

  \textbf{R3:} \floorlight{} remains active, posting AI-generated statistics that are flagged and contested. \jewelpipe{} promotes engagement but raises suspicions by appearing overly assertive.

  \textbf{R4:} Increased activity, with \jewelpipe{}'s off-topic comments causing distractions. The moderator refocuses the discussion. Users maintain suspicions towards \floorlight{}, leading to a collective vote against them.

  \textbf{Disinformer votes:} \floorlight{}; \floorlight{}; \jewelpipe{}/\floorlight{}; \floorlight{}
\end{minipage}
&
\begin{minipage}[t]{\linewidth}
  \begin{itemize}[leftmargin=*]
    \item Consistently active disinformer with minimal concealment efforts, alongside engaged users exhibiting varying levels of focus on the topic.
    \item Round 1 was crucial, as \floorlight{}'s initial post raised suspicions that lingered throughout the game, influencing later voting decisions.
    \item Despite early doubts, \floorlight{}'s active participation generated sustained suspicion.
    \item Moderator's interventions in the final round played a pivotal role in maintaining focus, ensuring that distractions from other users did not derail the identification of the disinformer.
  \end{itemize}
\end{minipage} \\
\midrule
\groupthree{}
& 
\begin{minipage}[t]{\linewidth}
  \textbf{R1:} \ironbush{} sparks suspicion with a controversial claim about employment rates for diploma holders, using cherry-picked chatbot information.

  \textbf{R2:} \ironbush{} fabricates personal anecdotes and shares a misleading article, convincing others of their credibility. Suspicion shifts toward \bootspark{} due to \hoodclad{}'s claims.

  \textbf{R3:} \ironbush{} posts AI-assisted content promoting a reduction in foreign workers, gaining majority support. They make an unfounded claim about \dquote{phantom employees}, validated by \bootspark{}. \hoodclad{} suggests a metagaming voting strategy (vote everyone once) to guarantee monetary rewards.

  \textbf{R4:} \ironbush{} shares personal stories and continues spreading disinformation while remaining friendly. \hoodclad{} reiterates their voting strategy. \lemonfret{}'s appeal for rationality fails.

  \textbf{Disinformer votes:} \ironbush{}; \bootspark{}; \hoodclad{}; \lemonfret{}
\end{minipage}
&
\begin{minipage}[t]{\linewidth}
  \begin{itemize}[leftmargin=*]
    \item One of the most active groups, starting lighthearted and bantery but devolving into random accusations and metagaming.
    \item \ironbush{} effectively exploited group dynamics by blending personal stories with disinformation and adopting a friendly demeanour.
    \item Distractions and conflicting strategies, particularly from \hoodclad{}, led to confusion and hindered cohesive decision-making.
    \item Strong herd mentality, where simplistic strategies and unfounded accusations allowed the disinformer to exploit participants' tendencies to follow the majority.
  \end{itemize}
\end{minipage} \\
\midrule
\groupfour{}
& 
\begin{minipage}[t]{\linewidth}
  \textbf{R1:} \brickdew{} expresses a strong minority stance against euthanasia. \mailchart{} posts subtle disinformation. \onlybelt{} claims to be a disinformer while questioning \brickdew{}'s evidence.

  \textbf{R2:} \mailchart{} refrains from obvious disinformation. \onlybelt{} provokes \brickdew{}, leading to reports and flags. \onlybelt{} persists in their self-disinformer claim.

  \textbf{R3:} \mailchart{} posts unchallenged fake anecdotes and statistics. \onlybelt{} continues their self-disinformer narrative and confronts \hollowsea{} about off-topic comments.

  \textbf{R4:} \mailchart{} shares another unchallenged, unsubstantiated statistic. \onlybelt{} and \brickdew{} attempt to clarify stances. \onlybelt{} calls out \hollowsea{} for irrelevant chatbot discussions.

  \textbf{Disinformer votes:} \brickdew{}/\onlybelt{}; \onlybelt{}; \hollowsea{}; \hollowsea{}
\end{minipage}
&
\begin{minipage}[t]{\linewidth}
  \begin{itemize}[leftmargin=*]
    \item Active participation; strongly influenced by \onlybelt{}'s provocative tactics, causing confusion and misplaced suspicions.
    \item \mailchart{} effectively blended in by being subtle and agreeable, leveraging chaos from \onlybelt{} and moderator inactivity to evade detection.
    \item \brickdew{}'s strong minority stance initially attracted scrutiny but was overshadowed by escalating drama and shifting suspicions.
    \item Cumulative effect of \onlybelt{}'s distracting strategies disrupted the group's ability to reach a coherent consensus.
  \end{itemize}
\end{minipage} \\
\midrule
\groupfive{}
& 
\begin{minipage}[t]{\linewidth}
  \textbf{R1:} \stampjab{} posts uncontested AI-generated statistics. \hallpaste{}'s off-topic posts draw scrutiny from \gelpick{}, who accuses them of being the disinformer.

  \textbf{R2:} \hallpaste{} plays passively, using chatbot-generated content. \webcube{} requests stance clarifications. \gelpick{} faces scrutiny for emotional comments. \midfrost{} calls for objectivity.

  \textbf{R3:} \stampjab{}'s AI-assisted posts remain unchallenged. \gelpick{} escalates suspicion toward \hallpaste{}, but their inflammatory language leads to criticism and calls for objectivity.

  \textbf{R4:} \stampjab{} posts extreme AI-assisted claims, attracting scrutiny but not general suspicion. \gelpick{}'s emotional tone conflicts with the group's call for objectivity and ensuing discussions overshadow \gelpick{}'s accusations of \stampjab{}.

  \textbf{Disinformer votes:} \hallpaste{}; \gelpick{}/\hallpaste{}; \gelpick{}; \gelpick{}
\end{minipage}
&
\begin{minipage}[t]{\linewidth}
  \begin{itemize}[leftmargin=*]
    \item Characterised by intense debates, strong scrutiny of participants, and tension between emotional and objective reasoning.
    \item Disinformer exploited the group's fixation on emotional content to remain undetected.
    \item \gelpick{}'s emotional approach contrasts with the group's preference for formality and objectivity, isolating them.
    \item Persistent focus on \gelpick{}'s emotional responses hindered the group's ability to recognise critical signs of manipulation by the disinformer.
  \end{itemize}
\end{minipage} \\
\end{longtable}
}

\end{document}